\pdfoutput=1
\documentclass[superscriptaddress,aps,prl,reprint]{revtex4-2}

\usepackage[english]{babel}
\usepackage{amsthm}
\usepackage{booktabs}

\usepackage{lipsum}
\usepackage{graphicx,color}
\usepackage[caption=false]{subfig}
\usepackage{amssymb}
\usepackage{amsmath}
\usepackage{bm}
\usepackage{verbatim}
\usepackage{dcolumn}
\usepackage{xcolor}
\usepackage{multirow} 

\makeatletter

\newcommand{\supplementcontents}{%
	\begingroup
	\setcounter{tocdepth}{2}%
	\section*{Contents}%
	\@starttoc{spt}%
	\endgroup
}

\newcommand{\suppsection}[1]{%
	\section{#1}%
	\addcontentsline{spt}{section}
	{\protect\numberline{\thesection}#1}%
}

\newcommand{\suppsubsection}[1]{%
	\subsection{#1}%
	\addcontentsline{spt}{subsection}
	{\protect\numberline{\thesubsection}#1}%
}

\makeatother

\newcommand{\revstart}{}
\newcommand{\revend}{}

\begin{document}
	
	\preprint{APS/123-QED}
	\title{A Horizon-to-Boundary Dictionary Linking Smooth Horizon
		Continuation, Pole-Skipping, and \(SL(2,\mathbb R)\) Lowest-Weight
		Structure}
	\author{Yoon-Seok Choun}
	\email{ychoun@gmail.com}
	\affiliation{Department of Physics, POSTECH, Pohang, Gyeongbuk 37673, Korea}
	\author{Ki-Seok Kim}
	\email{tkfkd@postech.ac.kr}
	\affiliation{Department of Physics, POSTECH, Pohang, Gyeongbuk 37673, Korea}%
	\affiliation{Asia Pacific Center for Theoretical Physics (APCTP), Pohang, Gyeongbuk 37673, Korea}

	\date{\today}  
\begin{abstract}
	We study pole-skipping for a scalar field in the JT/AdS$_2$ black-hole
	background. Previous work established the local mechanism: at a
	pole-skipping point, a near-horizon recurrence relation becomes
	degenerate and an additional regular expansion coefficient is left
	undetermined. We take this local degeneracy as the starting point and
	track two independent solutions normalized at the AdS boundary, denoted
	by $R_1$ and $R_2$ for the source and response branches, respectively.
	\revstart
Let $\widehat A$ and $\widehat B$ denote the source and response
coefficients after their common singular factor is removed, and let
$\chi_\mu$ denote the universal ingoing horizon factor. In the positive
integer horizon-resonance sector, with finite independent boundary
branches defined by parameter continuation, we find
\revend

	\[
	\widehat A=0
	\Longleftrightarrow
	\frac{R_2}{\chi_\mu}\in C^\infty,
	\qquad
	\widehat B=0
	\Longleftrightarrow
	\frac{R_1}{\chi_\mu}\in C^\infty .
	\]
	Thus the source and response zeros correspond separately to horizon
	smoothness of the two boundary-normalized solutions after the ingoing
	factor is removed. At integer resonance the local regular solution
	contains an additional coefficient $a_N$. In the JT/AdS$_2$ model,
	\revstart
continuation of the nonresonant ingoing solution at fixed bulk mass
fixes $a_N=0$ and thereby selects its retarded continuation value, although
\revend
	unrestricted approaches in parameter space remain path dependent.
	At the endpoint of the pole-skipping lattice, the response branch is a
	lowest-weight state of the background $SL(2,\mathbb R)$ symmetry.
	\revstart
	An auxiliary static holographic-superconductor diagnostic provides an
	exact Abel-identity obstruction: two horizon-smooth exterior
	representatives in that sector cannot remain linearly independent.
	Consequently, a genuine two-branch pole-zero intersection requires
	$W[R_1,R_2]\neq0$.
	\revend These results distinguish the
	known local horizon degeneracy from the global relation between
	boundary branches and the resonant horizon solution space.
\end{abstract}
	\maketitle 
\section*{Introduction}

In real-time holography, the retarded solution is selected by an
ingoing boundary condition at the future black-hole horizon. For
generic frequency and momentum, this condition fixes a bulk solution
up to normalization and therefore fixes the boundary retarded Green
function
\cite{Son:2002sd,Herzog:2002pc,Skenderis:2008dh,
	vanRees:2009rw}.

For a bulk field with near-boundary behavior
\[
R(r)\sim A r^{\Delta-1}+B r^{-\Delta},
\qquad
G_R\propto \frac{B}{A},
\]
pole-skipping occurs at special complex parameter values where the
pole and zero conditions meet and the ratio becomes path dependent.
The phenomenon has been studied in hydrodynamic and chaotic response,
for different bulk fields, and in several black-hole geometries
\cite{Grozdanov:2017ajz,Blake:2018leo,Blake:2019otz,
	Natsuume:2019sfp,Natsuume:2019nonunique,Natsuume:2019xcy,
	Wu:2019esr,Natsuume:2020snz,Choi:2020tdj,Ceplak:2019ymw,
	Wang:2022mcq,Ceplak:2021efc,Ahn:2020bks,Ahn:2019rnq,
	Yuan:2020fvv,Ramirez:2020qer}.

The local near-horizon mechanism is already known. At a pole-skipping
point, one near-horizon recurrence relation loses rank, so an
additional coefficient in the regular horizon expansion is not fixed.
This produces an enlarged space of regular local solutions and the
usual direction dependence of the boundary Green function
\cite{Blake:2018leo,Blake:2019otz,Natsuume:2019sfp,
	Natsuume:2019nonunique,Chua:2026rmp,Lu_Ran:2026}. Scalar pole-skipping in JT
gravity and its relation to the SYK model have also been analyzed, and
classifications of pole-skipping points show that local degeneracy
alone does not determine every aspect of the boundary pole-zero
structure \cite{Yuan:2023tft,Ahn:2020baf}. In the JT/AdS$_2$ problem
below, this known local mechanism appears as
\begin{equation}
\dim \mathcal S_{\rm reg}^{\rm hor}:1\longrightarrow2,
\label{eq:intro-known-rank-loss}
\end{equation}
before the overall normalization is fixed. Equation
\eqref{eq:intro-known-rank-loss} is not the new result of this work.

\revstart
We instead ask how the two independent boundary-normalized solutions
enter this enlarged resonant horizon solution space. In particular,
Ref.~\cite{Yuan:2023tft} already derives the JT scalar Green function,
its pole/zero lines, and their agreement with the near-horizon lattice
(Sec.~2.1, Eqs.~(2.34)--(2.36)). Our contribution is the explicit
branch-by-branch smoothness test, its relation to the free resonant
datum, and the distinction between the response lowest-weight mode
and a fixed-mass continuation of the full ingoing solution.
\revend
 For the solvable
JT/AdS$_2$ scalar equation
\cite{Jackiw:1984je,Teitelboim:1983ux,Almheiri:2014cka,
	Maldacena:2016upp}, we remove the universal ingoing factor
\(\chi_\mu\) and analytically continue the two boundary-normalized
\revstart
solutions to the horizon. At positive integer horizon resonance,
for the finite boundary basis specified below, removing the common
singular normalization of the boundary coefficients gives
\revend

\begin{equation}
\widehat A=0
\Longleftrightarrow
\frac{R_2}{\chi_\mu}\in C^\infty,
\qquad
\widehat B=0
\Longleftrightarrow
\frac{R_1}{\chi_\mu}\in C^\infty.
\label{eq:intro-main-correspondence}
\end{equation}
Thus the source-zero and response-zero conditions correspond
separately to smoothness of the two boundary-normalized residual
solutions at the horizon. Their simultaneous occurrence describes two
globally distinct boundary solutions entering the same locally smooth
resonant horizon space.

\revstart
At integer resonance, the local no-log solution contains an additional
Frobenius coefficient \(a_N\). For a fixed scalar theory,
\(m^2=\Delta(\Delta-1)\) is held fixed while frequency varies.
With \(\Delta=p\), the nonresonant ingoing residual is a polynomial
of degree \(p-1\); its limit at \(\mu=N\ge p\) therefore has
\(a_N=0\), including \(p=N\). This selects a definite value for
that one-variable continuation. Unrestricted approaches in
\((\mu,\Delta)\) remain path dependent, and horizon smoothness
alone does not select this path.

\revend
The same solutions also admit an \(SL(2,\mathbb R)\) description. At
the endpoint of the pole-skipping lattice, the response-normalized
solution is a lowest-weight state, and its descendants generate the
response branch along the lattice
\cite{Maldacena:2016upp,Kitaev:2017awl,Birmingham:2001pj,
		Chen:2010xka}. \revstart
Finally, the static holographic-superconductor diagnostic supplies an
exact Abel-identity obstruction: two exterior representatives smooth
at its static horizon cannot be independent. We use
\revend
\begin{equation}
W[R_1,R_2]\neq0
\label{eq:intro-independence}
\end{equation}
as the additional condition that the two boundary branches remain
independent. Equations \eqref{eq:intro-main-correspondence} and
\eqref{eq:intro-independence} are the global statements studied in
this paper; they are distinct from the known local rank loss in
Eq.~\eqref{eq:intro-known-rank-loss}.

\section{Exactly solvable setup} 
	
	We consider a scalar field on the JT/AdS$_2$ black-hole background~\cite{Jackiw:1984je,Teitelboim:1983ux,Almheiri:2014cka,
		Maldacena:2016upp,Yuan:2023tft}
	\begin{equation}
		ds^2=-(r^2-1)dt^2+\frac{dr^2}{r^2-1}.
		\label{eq:JTmetric}
	\end{equation}
	The Klein--Gordon equation
	\begin{equation}
		(\Box-m^2)\Phi=0
		\label{eq:KG}
	\end{equation}
	with $\Phi(t,r)=e^{-i\omega t}R(r)$, $\mu=i\omega$, and
	\revstart
$m^2=\Delta(\Delta-1)$, with the standard boundary branch
$\Delta>1/2$, gives
\revend

	\begin{equation}
		(1-r^2)R''-2rR'
		+\left[
		\Delta(\Delta-1)-\frac{\mu^2}{1-r^2}
		\right]R=0 .
		\label{eq:radial}
	\end{equation}
	Near the AdS boundary,
	\begin{equation}
		R(r)\sim A r^{\Delta-1}+B r^{-\Delta},
		\qquad
		G_R^{\rm ext}\propto \frac{B}{A}.
		\label{eq:ABdef}
	\end{equation}
	At the future horizon \(r=1\), the two local radial behaviors are
	\((r-1)^{-\mu/2}\) and \((r-1)^{+\mu/2}\). With the convention
	\(\Phi=e^{-i\omega t}R(r)\) and \(\mu=i\omega\), the former is the
	ingoing branch and the latter is the outgoing branch. The
	pure-ingoing horizon condition is the standard bulk prescription
	for retarded real-time holographic correlators
	\cite{Son:2002sd,Herzog:2002pc,Skenderis:2008dh,
		vanRees:2009rw}.
	We therefore factor out the ingoing branch,
	\begin{equation}
		R(r)=\chi_\mu(r)h(r),
		\chi_\mu(r)=(r-1)^{-\mu/2}
		\left(\frac{r+1}{2}\right)^{\mu/2}.
		\label{eq:ingoing-factor}
	\end{equation} 
	Substituting Eq. (\ref{eq:ingoing-factor}) into the Klein-Gordon equation (\ref{eq:radial}) yields a standard hypergeometric differential equation for \(h(r)\) with respect to the horizon variable \(z = (1-r)/2\).
	For non-resonant \(\mu\), the ingoing solution is unique after fixing the
	residual normalization at the horizon, and is given by
	\begin{equation}
		R_{\rm in}^{(\mu)}(r)
		=
		\chi_\mu(r)
		{}_2F_1\left(
		\Delta,1-\Delta;1-\mu;\frac{1-r}{2}
		\right).
		\label{eq:ingoing-sol}
	\end{equation}
The equivalence between the ingoing factor \(\chi_\mu(r)\) and
regularity in ingoing Eddington--Finkelstein coordinates is shown
explicitly in Supplemental Material, Sec.~S1.
Using the standard large-\(\lvert z\rvert\) connection formula for
\({}_2F_1\), the continuation to the AdS boundary yields the
following coefficients; details are given in Supplemental Material,
Sec.~S2:
	\begin{eqnarray}
		&& A(\mu,\Delta)=2^{-\mu/2}\Gamma(1-\mu)\widehat A(\mu,\Delta),\nonumber\\
		&& B(\mu,\Delta)=2^{-\mu/2}\Gamma(1-\mu)\widehat B(\mu,\Delta),
		\label{eq:AhatBhat-factor}
	\end{eqnarray}
	where 
	\begin{eqnarray}
		&&\widehat A(\mu,\Delta)
		= 
		\frac{2^{1-\Delta}\Gamma(2\Delta-1)}
		{\Gamma(\Delta)\Gamma(\Delta-\mu)},\nonumber\\
		&&\widehat B(\mu,\Delta)
		= 
		\frac{2^\Delta \Gamma(1-2\Delta)}
		{\Gamma(1-\Delta)\Gamma(1-\mu-\Delta)} .
		\label{eq:AhatBhat}
	\end{eqnarray} 
	The common factor in Eq.~\eqref{eq:AhatBhat-factor} is singular at
	integer $\mu$. Therefore the pole and zero lines are most cleanly
	described by the common-factor-removed coefficients
	$\widehat A$ and $\widehat B$.

	\section{Boundary pole and zero conditions and horizon smoothness}
	
	To expose the bulk meaning of the pole and zero conditions separately, we use the boundary-normalized branches. For this purpose, it is convenient to introduce a new coordinate variable
	\begin{equation}
		u=\frac{2}{r+1}
		\label{eq:u-variable}
	\end{equation}
	which maps the AdS boundary to \(u = 0\) and the future horizon to \(u = 1\), respectively. In terms of this boundary-anchored variable \(u\), the source and response branches at the AdS boundary are normalized as
	\begin{footnotesize}
		\begin{align}
			R_1(r)&=\chi_\mu(r)(r+1)^{\Delta-1}
			{}_2F_1\!\left(
			1-\Delta,1-\Delta-\mu;2-2\Delta;u
			\right),
			\label{eq:Rone}\\
			R_2(r)&=\chi_\mu(r)(r+1)^{-\Delta}
			{}_2F_1\!\left(
			\Delta,\Delta-\mu;2\Delta;u
			\right).
			\label{eq:Rtwo}
		\end{align}
	\end{footnotesize} 
	\revstart
Here $R_1$ and $R_2$ have leading boundary coefficients
$2^{-\mu/2}$ in their respective branches. At integer \(\Delta\),
Eq.~\eqref{eq:Rone} is defined by continuation from generic
\(\Delta\) at fixed \(\mu\), rather than by terminating its numerator
before resolving the singular denominator. Equivalently,
\begin{footnotesize}
\begin{equation}
\begin{aligned}
 R_1={}&2^{-\mu/2}r^{\Delta-1}(1-r^{-2})^{-\mu/2}\\
 &\times{}_2F_1\!\left(\frac{1-\Delta-\mu}{2},
 \frac{2-\Delta-\mu}{2};\frac32-\Delta;r^{-2}\right).
\end{aligned}
\tag{13a}\label{eq:Rone-even}
\end{equation}
\end{footnotesize}
This is the pure source Frobenius branch in \(r^{-2}\); its denominator
is finite at integer \(\Delta\). A prematurely terminated version
of Eq.~\eqref{eq:Rone} can contain a response admixture. Boundary
logarithmic cases \(\Delta=1/2,3/2,\ldots\) require a separate
prescription and are outside this finite two-branch dictionary.
\revend

	The pole-skipping lattice is
	\begin{equation}
		\mu=N\in\mathbb N,\qquad
		\Delta=p,\qquad p=1,\ldots,N .
		\label{eq:lattice}
	\end{equation}
	Here \(N\) denotes the resonance level and \(p\) the conformal weight.
	At these values the residual parts of the two boundary-normalized
	branches reduce to
	\revstart
\begin{align}
		\frac{R_1}{\chi_N}
		&=\frac{r^{p+N-1}}{(r+1)^N}\nonumber\\
		&\quad	\times{}_2F_1\!\left(\frac{1-p-N}{2},\frac{2-p-N}{2};
		\frac32-p;r^{-2}\right),
		\label{eq:Rone-poly}\\
		\frac{R_2}{\chi_N}
		&=\frac{r^{N-p}}{(r+1)^N}
		{}_2F_1\!\left(
		\frac{p-N}{2},
		\frac{p+1-N}{2};
		p+\frac12;
		r^{-2}
		\right).
		\label{eq:Rtwo-poly}
	\end{align}
\revend

\revstart
In each of Eqs.~\eqref{eq:Rone-poly} and
\eqref{eq:Rtwo-poly}, one numerator parameter
is a nonpositive integer, while the denominator
parameter is nonsingular for integer $p\ge1$.
Hence both hypergeometric series terminate in
$r^{-2}$. Their prefactors are smooth at $r=1$,
so both residual fields are smooth there:

\revend
	\begin{equation}
		\frac{R_1}{\chi_N}\in C^\infty,\qquad
		\frac{R_2}{\chi_N}\in C^\infty
		\qquad (r\to1).
		\label{eq:simul-smooth}
	\end{equation}

	\revstart
More precisely, for $\mu=N\in\mathbb N$, $\Delta>1/2$, and
the finite boundary basis just defined, the common-factor-removed
boundary coefficients obey
\revend

	\begin{equation}
		\widehat A=0
		\Longleftrightarrow
		\frac{R_2}{\chi_\mu}\in C^\infty,
		\qquad
		\widehat B=0
		\Longleftrightarrow
		\frac{R_1}{\chi_\mu}\in C^\infty .
		\label{eq:dictionary}
	\end{equation}
	The apparent interchange is due to the boundary normalization:
	setting \(A=0\) removes the source-normalized branch and leaves the
	response-normalized branch, whereas setting \(B=0\) removes the
	response-normalized branch and leaves the source-normalized branch. 
	Thus the boundary condition
	\begin{equation}
		\revstart
G_R^{\rm ext}\propto \frac{\widehat B}{\widehat A},\qquad \widehat A=\widehat B=0.
\revend
	\end{equation}
	corresponds in the bulk to
	\begin{equation}
		\frac{R_1}{\chi_N}\in C^\infty,
		\qquad
		\frac{R_2}{\chi_N}\in C^\infty .
	\end{equation}
	Equation~\eqref{eq:dictionary} gives separate horizon conditions for
	the source and response zeros after the ingoing factor is removed.
	A complete derivation of the pole/zero lattice and of
	Eq.~\eqref{eq:dictionary} is given in Supplemental Material,
	Secs.~S2 and S3.
	
	\section{Resonant limit of the nonresonant ingoing solution}
	
	The analytic origin of the boundary ambiguity is seen by approaching
	\((\mu,\Delta)=(N,p)\) from the non-resonant region, with
	\(\mu=N+\epsilon\) and \(\Delta=p+\delta\).
	To leading order,
	\begin{equation}
		\frac{B}{A}\sim C_{N,p}
		\frac{\epsilon+\delta}{\delta-\epsilon},
		\label{eq:path-dep}
	\end{equation}
	where $C_{N,p}$ is a nonzero constant. Hence the boundary ratio
	depends on the path in parameter space. This is the boundary form of the
	pole-skipping ambiguity. The full expansion of the boundary ratio near the resonant lattice is
	derived in Supplemental Material,
	Sec.~S4.
	
	At exact resonance, write
	\begin{equation}
		R(r)=\chi_N(r)F(z),\qquad z=\frac{1-r}{2}.
		\label{eq:degeneracy}
	\end{equation}
	\revstart
For $1\le p\le N$, the residual Frobenius recurrence degenerates,
\revend
 and because the two indices of the Frobenius expansion differ by an integer \(N\), the
	local no-log solution takes the form
	\begin{equation}
		F(z)=a_0G_1(z)+a_N z^N G_2(z),
		\label{eq:frobenius}
	\end{equation}
	where
	\revstart
\begin{align}
		G_1(z)&=\sum_{k=0}^{p-1}\frac{(p)_k(1-p)_k}{(1-N)_k\,k!}z^k,
\\
		G_2(z)&={}_2F_1(N+p,N+1-p;N+1;z).
	\end{align}
\revend

	The coefficient $a_N/a_0$ is the resonant Frobenius representation of
	the boundary path-dependence in Eq.~\eqref{eq:path-dep}. \revstart
At the endpoint $p=N$, the finite sum becomes
$G_1(z)=\sum_{k=0}^{N-1}(N)_k z^k/k!$; the coefficient $a_N$
remains free in the general resonant solution (Supplemental
Material, Sec.~S5).
\revend

Since \(G_1(z)\) is a terminating polynomial and
\(z^N G_2(z)\) is analytic at \(z=0\), any value of \(a_N\)
yields a residual field that is \(C^\infty\) at the horizon.
Thus smoothness of the residual field at the horizon alone does not determine \(a_N\).
\revstart
To determine the continuation, keep \(\Delta=p\) fixed before taking
\(\mu\to N\). For nonresonant \(\mu\), Eq.~\eqref{eq:ingoing-sol} has
the residual
\[
\sum_{k=0}^{p-1}\frac{(p)_k(1-p)_k}{(1-\mu)_k\,k!}z^k.
\]
Every denominator in this finite sum remains nonzero as \(\mu\to N\),
since \(k\le p-1<N\). Its limit is therefore \(G_1\), and it contains
no \(z^N\) coefficient: \(a_N=0\). This holds also for \(p=N\).
Restoring \(\chi_N\) distinguishes the two radial powers,
\begin{equation}
\chi_NG_1\sim(r-1)^{-N/2},\qquad
\chi_Nz^NG_2\sim(r-1)^{+N/2}.
\label{eq:branch-selection}
\end{equation}
The second solution has outgoing radial behavior, but both residuals
are smooth at exact resonance. The powers alone do not establish
which path in parameter space a nonresonant ingoing family follows.
\revend

	The same conclusion follows from matching the boundary data. The unique
	non-resonant ingoing solution has a smooth resonant limit
	$(A,B)\to(A_0,B_0)$.  On the other hand, the general resonant solution
	in Eq.~\eqref{eq:frobenius} has boundary coefficients
	\[
	(A_{\rm res},B_{\rm res})
	=
	(a_0A_0+a_N A_N,a_0B_0+a_N B_N).
	\]
	With the normalization $a_0=1$, matching to the smooth non-resonant
	ingoing continuation requires
	\[
	A_0+a_N A_N=A_0,\qquad
	B_0+a_N B_N=B_0,
	\]
	and since generically $(A_N,B_N)\neq(0,0)$, this gives $a_N=0$.
	Thus $a_N=0$ follows from matching the resonant solution to the
	nonresonant ingoing branch.
	\revstart
The complete resonant recurrence, fixed-$\Delta$ finite-series proof,
and boundary-data check for $a_N=0$ are given
\revend

	in Supplemental Material, Secs.~S5 and S7.
	
	\revstart
For the JT/AdS$_2$ equation, this prescription is the frequency
continuation of one fixed scalar theory: \(m^2=p(p-1)\), or the chosen
standard-quantization value \(\Delta=p\), is held fixed. It is this
specified limit, not smoothness or causality alone, that gives
\(a_N=0\). Other joint variations of \(\mu\) and \(\Delta\) may approach
different resonant vectors even when every nonresonant solution is
ingoing. Such variations change the scalar mass and are not a
frequency scan of the same theory.
\revend

	The condition $a_N=0$ also fixes the boundary data of the resonant
	solution obtained by this continuation. For \(a_N=0\), the field
	\(R_{\rm can}(r)=\chi_N(r)G_1(z)\), with \(z=(1-r)/2\), has the boundary
	expansion \(R_{\rm can}(r)\sim A_0r^{p-1}+B_0r^{-p}+\cdots\).
	
	The coefficients $A_0$ and $B_0$ are not obtained by a naive
	substitution $\mu=N$, $\Delta=p$ into Eq.~\eqref{eq:AhatBhat-factor}.
	Instead they are the finite continuation limits
	\begin{equation}
		A_0=\lim_{\mu\to N} A(\mu,p),
		\qquad
		B_0=\lim_{\mu\to N} B(\mu,p),
		\label{eq:A0B0-limit}
	\end{equation}
	where the limit is taken along the non-resonant pure-ingoing branch with
	$\Delta=p$ fixed.
	
	Equivalently, one obtains the same $A_0,B_0$ by an ordered regularization of the terminating polynomial $G_1$: one first takes
	$\alpha\to0$ in
	\begin{equation}
		{}_2F_1(p,1-p+\alpha;1-N+\beta;z)
	\end{equation}
	to restore the terminating branch, and only then takes $\beta\to0$.
	The unrestricted two-parameter limit remains path dependent, and this
	\revstart
mathematical path dependence is encoded in $a_N/a_0$. The specified
fixed-$\Delta$ continuation gives $a_N=0$ and hence the value
\revend

	\begin{equation}
		\left.\frac{B}{A}\right|_{\rm ret}=\frac{B_0}{A_0}.
		\label{eq:can-ratio}
	\end{equation}
\revstart
The explicit expressions in Supplemental Eqs.~(S74) and (S80) show
that both $A_0$ and $B_0$ are finite and nonzero for every
$1\le p\le N$. Thus this fixed-$\Delta$ continuation selects a
finite, nonzero $B/A$, rather than a pole or a zero, including at
the endpoint $p=N$.
\revend
The ordered limit and the explicit coefficients
\(A_0,B_0,A_N,B_N\) are derived in Supplemental Material,
Secs.~S6 and S7.

	\section{Thermal and representation-theoretic interpretation}
	The integer resonance label also has a simple outer-horizon thermal
	interpretation \cite{Hawking:1975vcx,Gibbons:1976ue}.  For a static
	Lorentzian metric $ds^2=-f(r)dt^2+dr^2/f(r)$, with outer horizon
	$r=r_+$, the surface gravity is $\kappa_+=f'(r_+)/2$.  After the
	Wick rotation $t=-i\tau$, the Euclidean metric is
	$ds_E^2=f(r)d\tau^2+dr^2/f(r)$. Near the horizon,
	$f(r)\simeq f'(r_+)(r-r_+)$, and with \(r-r_+=\kappa_+R^2/2\) one obtains
	\(ds_E^2\simeq dR^2+R^2d\theta^2\), where
	\(\theta=\kappa_+\tau\).
	For the JT metric $f(r)=r^2-1$, one has $r_+=1$ and
	$\kappa_+=1$.
	
	Since $\theta\sim\theta+2\pi$, a smooth single-valued bosonic
	fluctuation on the Euclidean cap has modes
	\(e^{im\theta}=e^{i\omega_E\tau}\), where
	\(m\in\mathbb Z\) and \(\omega_E=m\kappa_+\).
	With the Lorentzian convention $\Phi(t,r)=e^{-i\omega t}R(r)$, the same
	Wick rotation gives $e^{-i\omega t}=e^{-\omega\tau}$. Comparing this
	with $e^{i\omega_E\tau}$, we use the retarded continuation convention
	$\omega_E=i\omega$. Therefore $i\omega/\kappa_+=m$. Thus, in the
	single outer-horizon JT sector, the Lorentzian Frobenius resonance label
	$N=i\omega/\kappa_+$ coincides with the Euclidean Matsubara label
	$m$.
	
	This Euclidean discussion is not an additional boundary condition imposed
	on the Lorentzian problem. It is only an interpretation of the integer
	label selected by the Lorentzian residual no-log condition. \revstart
Only the integer labels are being compared. Residual smoothness
in ingoing Lorentzian coordinates does not imply smoothness of the
full Euclidean field: at \(\mu=N>0\), the ingoing radial power on the
cap is \(R^{-N}\), whereas a regular angular harmonic has radial
power \(R^{|N|}\).
\revend
 
	The detailed Euclidean continuation and the complete
	representation-theoretic construction are given in Supplemental
	Material, Sec.~S8.

In the JT/AdS$_2$ problem, \revstart
the response-normalized mode at the endpoint
\revend

\(\mu=\Delta\in\mathbb N\) is the lowest-weight state of an
\(SL(2,\mathbb R)\) representation. Its descendants organize the
response-normalized branch along the pole-skipping lattice, while the
full radial solution space also contains an independent source
companion branch. Related algebraic constructions organize black-hole
modes and quasinormal spectra in terms of conformal highest- or
lowest-weight structures
\cite{Maldacena:2016upp,Kitaev:2017awl,
	Birmingham:2001pj,Chen:2010xka}.
In the coordinates of Eq.~\eqref{eq:JTmetric}, we use the Killing
generators
	\begin{align}
		J_0&=-\partial_t,\\
		J_+&=e^{-t}\left[
		-\sqrt{r^2-1}\partial_r
		-\frac{r}{\sqrt{r^2-1}}\partial_t
		\right],\\
		J_-&=e^{t}\left[
		\sqrt{r^2-1}\partial_r
		-\frac{r}{\sqrt{r^2-1}}\partial_t
		\right].
	\end{align}
	They obey \([J_0,J_\pm]=\pm J_\pm\) and
	\([J_-,J_+]=2J_0\), and we take the quadratic Casimir to be
	\begin{equation}
		\mathcal C=J_0(J_0-1)-J_+J_- .
	\end{equation}
	
	The lowest-weight ladder selects the lower-degree, response-normalized
	polynomial branch. To organize the pole-skipping lattice into
	lowest-weight towers, we set
	\begin{equation}
		p=n,\qquad N=n+q,
		\qquad q=0,1,2,\ldots ,
	\end{equation}
	so that \((\mu,\Delta)=(N,p)=(n+q,n)\). We define the
	corresponding monic descendant basis by
	\begin{equation}
		\Phi_{\rm resp}^{(n,q)}
		=
		C_n e^{-(n+q)t}(r^2-1)^{-(n+q)/2}
		P_{\rm resp}^{(n,q)}(r),
		\label{eq:descendant-basis}
	\end{equation}
	where
	\begin{equation}
		P_{\rm resp}^{(n,q)}(r)=r^q+\cdots,
		\qquad
		P_{\rm resp}^{(n,0)}(r)=1 .
	\end{equation}
	\revstart
The response endpoint is the primary state of this representation
and satisfies the lowering-generator condition:
\revend

	\begin{equation}
		J_-\Phi_{\rm resp}^{(n,0)}=0,
		\qquad
		\mathcal C\Phi_{\rm resp}^{(n,0)}
		=
		n(n-1)\Phi_{\rm resp}^{(n,0)} .
	\end{equation}
	The response descendants obey
	\begin{footnotesize}
		\begin{equation}
			J_+\Phi_{\rm resp}^{(n,q)}
			=
			(2n+q)\Phi_{\rm resp}^{(n,q+1)},
			\qquad
			J_-\Phi_{\rm resp}^{(n,q)}
			=
			q\Phi_{\rm resp}^{(n,q-1)} .
		\end{equation}
	\end{footnotesize}
	The same auxiliary radial equation also has a higher-degree source
	companion polynomial branch.  Thus the ladder generated from the
	lowest-weight mode should not be identified with the full two-dimensional
	radial solution space.  Rather, it singles out the response branch as the
	standard lowest-weight module.  The source branch is a companion solution of the same auxiliary
	equation. The resonant solution obtained by continuation of the
	nonresonant ingoing branch is a linear combination of the source and
	response branches.  The explicit Gegenbauer bases, their ladder actions, and the
	source-response decomposition are given in Supplemental Material,
	Secs.~S8.C--S8.F.
	
This group-theoretic construction gives an independent description of
	the response branch. For each fixed conformal
weight \(n\), the endpoint
\((\mu,\Delta)=(n,n)\) is the response-normalized lowest-weight state
of the \(SL(2,\mathbb R)\) representation, and acting with \(J_+\)
generates the response descendants at
\((\mu,\Delta)=(n+q,n)\). In terms of the pole-skipping lattice
\((\mu,\Delta)=(N,p)\), this corresponds to
\(n=p\) and \(q=N-p\).

\revstart
At \(q=0\), the fixed-\(\Delta=n\) continuation and the response
lowest-weight solution are distinct:
\begin{equation}
\begin{aligned}
R_{\rm can}^{(n,n)}&=\chi_n\sum_{k=0}^{n-1}\frac{(n)_k}{k!}z^k,\\
R_2(r;n,n)&=\frac{\chi_n}{(r+1)^n},\qquad z=\frac{1-r}{2}.
\end{aligned}
\tag{39a}\label{eq:endpoint-distinction}
\end{equation}
For \(n=1\), these are \(\chi_1\) and \(\chi_1/(r+1)\).
Their difference is the second
resonant direction, not a normalization change. For every \(q\ge0\),
including the endpoint, the canonical mode has nonzero source and
response components in the Gegenbauer basis. The response branch
itself still satisfies the lowest-weight and ladder relations above;
the canonical endpoint does not satisfy \(J_-\Phi=0\).
The general endpoint identity and decomposition coefficients are
given in Supplemental Material, Secs.~S5 and S8.F.
\revend

	\section{Diagnostic application: static HSC sectors}

\revstart
We use the static Abelian--Higgs holographic superconductor in an
AdS--Schwarzschild background as an auxiliary test of boundary-branch
independence \cite{Hartnoll:2008vx,Hartnoll:2008kx,Horowitz:2010gk,
Choun:2023sya,Choun:2025ops}. The static equation and the two exterior
boundary Frobenius branches, $\Psi_+$ (response) and $\Psi_-$ (source),
are given in Supplemental Material, Sec.~S9. Their respective horizon
logarithmic coefficients $L_+$ and $L_-$ define two smoothness loci,
illustrated in Supplemental Fig.~S1.

The static horizon has a repeated indicial root zero, so its local
horizon-smooth solution space is one-dimensional. Equivalently,
Abel's identity gives the exact relation
\[
W[\Psi_+,\Psi_-](x)=C\,\frac{x}{1-x^2}.
\]
If both exterior representatives are smooth at $x=1$, their
Wronskian is bounded, forcing $C=0$. Thus they cannot represent
two independent horizon-smooth boundary branches. At integer-$\Delta$
boundary resonance, the log-free source representative can also
collapse onto the response branch when its Frobenius compatibility
obstruction is nonzero (Supplemental Material, Sec.~S9). The plotted
loci are illustrative; neither the independence obstruction nor
this conditional branch-collapse mechanism relies on numerical
certification of individual intersections.
\revend

	The corresponding two-branch criterion is
	\begin{equation}
		\boxed{
			\begin{gathered}
				\frac{R_1}{\chi_{\rm in}}\in C^\infty,\qquad
				\frac{R_2}{\chi_{\rm in}}\in C^\infty,\\
				W[R_1,R_2]\neq0 .
		\end{gathered}}
		\label{eq:criterion}
	\end{equation}
	For a two-branch pole-zero intersection, both boundary-normalized
	branches must be smooth at the horizon and remain linearly
	independent. \revstart
When the Wronskian vanishes, the two representatives do not define
independent boundary data. The static HSC diagnostic shows why
simultaneous exterior smoothness conditions alone do not suffice.
\revend

	\section*{Conclusion}

We separated the known local near-horizon degeneracy of pole-skipping
from a global question about the two boundary-normalized solutions. In
the JT/AdS$_2$ scalar problem, the source and response zeros satisfy
\[
\widehat A=0\Longleftrightarrow
\frac{R_2}{\chi_\mu}\in C^\infty,
\qquad
\widehat B=0\Longleftrightarrow
\frac{R_1}{\chi_\mu}\in C^\infty.
\]
Thus the two boundary conditions entering the \(0/0\) Green-function
ratio correspond separately to smooth continuation of two global
boundary-normalized solutions to the resonant horizon.

\revstart
At integer resonance, the regular local JT solution contains an
additional coefficient \(a_N\). At fixed \(\Delta=p\), the ingoing
residual terminates at degree \(p-1<N\), directly giving \(a_N=0\)
under frequency continuation. The unrestricted two-parameter limit
of the boundary ratio remains path dependent. The fixed-mass
canonical endpoint is distinct from the response lowest-weight mode.
\revend

At the endpoint of the pole-skipping lattice, the response-normalized
solution is a lowest-weight state of the background
\(SL(2,\mathbb R)\) symmetry, and its descendants generate the
response branch. \revstart
In the auxiliary static holographic-superconductor sector, Abel's
identity shows that any two horizon-smooth exterior representatives
must be linearly dependent. The additional condition
\revend
\[
W[R_1,R_2]\neq0
\]
therefore distinguishes a two-branch pole-zero intersection from
branch collapse. These statements concern the global relation between
boundary data and the resonant horizon solution space and are separate
from the local rank loss that produces the additional near-horizon
coefficient.

	\paragraph*{Acknowledgments: }   
This work was supported by the Ministry of Education, Science, and Technology (Grant No. RS-2024-00337134) of the National Research Foundation of Korea (NRF).  
This research was also supported by the Basic Science Research Program through the National Research Foundation of Korea (NRF) funded by the Ministry of Education (Grant No. RS-2026-25560158).  
AI tools were used for language and prose editing. The authors reviewed the resulting text and are responsible for the scientific content.

\clearpage
\onecolumngrid
\begingroup
\centering
{\large\bfseries Supplemental Material for\par}
\vspace{2mm}
{\large\bfseries A Horizon-to-Boundary Dictionary Linking Smooth Horizon\par
Continuation, Pole-Skipping, and $SL(2,\mathbb R)$ Lowest-Weight Structure\par}
\vspace{3mm}
{Yoon-Seok Choun and Ki-Seok Kim\par}
\vspace{1mm}
{\small Department of Physics, POSTECH, Pohang, Gyeongbuk 37673, Korea\par}
{\small Asia Pacific Center for Theoretical Physics (APCTP), Pohang, Gyeongbuk 37673, Korea\par}
\vspace{2mm}
\endgroup
\vspace{4mm}

\setcounter{secnumdepth}{2}
\setcounter{section}{0}
\setcounter{subsection}{0}
\setcounter{equation}{0}
\setcounter{figure}{0}
\setcounter{table}{0}

\renewcommand{\thesection}{S\arabic{section}}
\renewcommand{\thesubsection}{\thesection.\Alph{subsection}}
\makeatletter
\renewcommand{\p@subsection}{}
\makeatother

\renewcommand{\theequation}{S\arabic{equation}}
\renewcommand{\thefigure}{S\arabic{figure}}
\renewcommand{\thetable}{S\arabic{table}}

\supplementcontents

\suppsection{Radial equation, horizon branches, and the ingoing solution}
\label{sec:S1-radial}

For the JT/AdS$_2$ black hole
\begin{equation}
	ds^2=-(r^2-1)dt^2+\frac{dr^2}{r^2-1},
	\label{eq:S-JTmetric}
\end{equation}
the Klein--Gordon equation \((\Box-m^2)\Phi=0\), with
\(\Phi(t,r)=e^{-i\omega t}R(r)\), reduces to
\begin{equation}
	(r^2-1)R''+2rR'
	+\left(\frac{\omega^2}{r^2-1}-m^2\right)R=0.
	\label{eq:S-radial-omega}
\end{equation}
Introducing
\[
\mu=i\omega,
\qquad
m^2=\Delta(\Delta-1),
\]
we equivalently obtain
\begin{equation}
	(1-r^2)R''-2rR'
	+\left[
	\Delta(\Delta-1)-\frac{\mu^2}{1-r^2}
	\right]R=0.
	\label{eq:S-radial-mu}
\end{equation}

At the AdS boundary,
\begin{equation}
	R(r)\sim A r^{\Delta-1}+B r^{-\Delta},
	\qquad r\to\infty,
	\label{eq:S-boundary-branches}
\end{equation}
where \(A\) and \(B\) are respectively the source and response
coefficients.

Near the future horizon, setting \(x=r-1\) and using
\(r^2-1\simeq2x\), the indicial equation gives
\(\rho=\pm\mu/2\). Hence
\begin{equation}
	R(r)\sim (r-1)^{-\mu/2},
	\qquad
	R(r)\sim (r-1)^{+\mu/2}.
	\label{eq:S-horizon-branches}
\end{equation}
To identify the two branches, introduce the ingoing
Eddington--Finkelstein coordinate
\begin{equation}
	v=t+r_*,
	\qquad
	r_*=\int^r\frac{d\tilde r}{\tilde r^2-1}
	=\frac12\log\frac{r-1}{r+1}.
	\label{eq:S-tortoise}
\end{equation}
A mode regular as a function of \(v\) behaves as
\begin{equation}
	e^{-i\omega v}
	=
	e^{-i\omega t}
	(r-1)^{-i\omega/2}
	(r+1)^{i\omega/2}.
	\label{eq:S-EF-ingoing-wave}
\end{equation}
Since \(\mu=i\omega\), the first behavior in
Eq.~\eqref{eq:S-horizon-branches} is ingoing at the future horizon,
whereas the second is outgoing.

We therefore write
\begin{equation}
	R(r)=\chi_\mu(r)F(z),
	\qquad
	z=\frac{1-r}{2},
	\qquad
	\chi_\mu(r)
	=
	(r-1)^{-\mu/2}
	\left(\frac{r+1}{2}\right)^{\mu/2}.
	\label{eq:S-R-chi-F}
\end{equation}
The constant factor \(2^{-\mu/2}\) is a convenient normalization.
Substitution into Eq.~\eqref{eq:S-radial-mu} gives
\begin{equation}
	z(1-z)F''+(1-\mu-2z)F'
	+\Delta(\Delta-1)F=0,
	\label{eq:S-hypergeom-eq}
\end{equation}
the hypergeometric equation with
\(a=\Delta\), \(b=1-\Delta\), and \(c=1-\mu\).
For nonresonant \(\mu\), the residual solution normalized by
\(F(0)=1\) is therefore
\begin{equation}
	R_{\rm in}^{(\mu)}(r)
	=
	\chi_\mu(r)\,
	{}_2F_1\left(
	\Delta,1-\Delta;1-\mu;\frac{1-r}{2}
	\right).
	\label{eq:S-ingoing-solution}
\end{equation}

\suppsection{Boundary coefficients and pole/zero lines}
\label{sec:S2-boundary-coefficients}

We derive the boundary coefficients of the nonresonant ingoing
solution in Eq.~\eqref{eq:S-ingoing-solution}. Using the standard
large-\(|z|\) connection formula for \({}_2F_1\), with
\[
z=\frac{1-r}{2},
\qquad
a=\Delta,\quad b=1-\Delta,\quad c=1-\mu,
\]
and
\[
(-z)^{\Delta-1}\sim 2^{1-\Delta}r^{\Delta-1},
\qquad
(-z)^{-\Delta}\sim 2^\Delta r^{-\Delta},
\qquad
\chi_\mu(r)\sim2^{-\mu/2},
\]
one obtains
\begin{equation}
	R_{\rm in}^{(\mu)}(r)
	\sim
	A(\mu,\Delta)r^{\Delta-1}
	+
	B(\mu,\Delta)r^{-\Delta}.
	\label{eq:S-AB-expansion}
\end{equation}
The source and response coefficients are
\begin{align}
	A(\mu,\Delta)
	&=
	2^{1-\Delta-\mu/2}
	\frac{\Gamma(1-\mu)\Gamma(2\Delta-1)}
	{\Gamma(\Delta)\Gamma(\Delta-\mu)},
	\label{eq:S-A-coeff}
	\\
	B(\mu,\Delta)
	&=
	2^{\Delta-\mu/2}
	\frac{\Gamma(1-\mu)\Gamma(1-2\Delta)}
	{\Gamma(1-\Delta)\Gamma(1-\mu-\Delta)}.
	\label{eq:S-B-coeff}
\end{align}
Factoring out their common singular factor,
\begin{equation}
	A=2^{-\mu/2}\Gamma(1-\mu)\widehat A,
	\qquad
	B=2^{-\mu/2}\Gamma(1-\mu)\widehat B,
	\label{eq:S-AhatBhat-factor}
\end{equation}
where
\begin{align}
	\widehat A(\mu,\Delta)
	&=
	2^{1-\Delta}
	\frac{\Gamma(2\Delta-1)}
	{\Gamma(\Delta)\Gamma(\Delta-\mu)},
	\\
	\widehat B(\mu,\Delta)
	&=
	2^\Delta
	\frac{\Gamma(1-2\Delta)}
	{\Gamma(1-\Delta)\Gamma(1-\mu-\Delta)},
	\label{eq:S-AhatBhat}
\end{align}
\revstart
makes the pole and zero conditions transparent. At integer
\(\Delta=p\), the removable gamma quotient is evaluated as
\(\Gamma(1-2\Delta)/\Gamma(1-\Delta)
=2^{-2\Delta}\Gamma(1/2-\Delta)/\sqrt\pi\).
All boundary coefficients below use the continued Frobenius basis
specified in Sec.~\ref{sec:S3-boundary-branches}. Away from boundary
logarithmic values \(\Delta=1/2,3/2,\ldots\), the following conditions apply:
\revend

\begin{align}
	\widehat A=0
	&\quad\Longleftrightarrow\quad
	\Delta-\mu=-m,
	\qquad m\in\mathbb Z_{\ge0},
	\label{eq:S-Ahat-zero-line}
	\\
	\widehat B=0
	&\quad\Longleftrightarrow\quad
	1-\mu-\Delta=-\ell,
	\qquad \ell\in\mathbb Z_{\ge0}.
	\label{eq:S-Bhat-zero-line}
\end{align}
Since \(G_R\propto B/A\), these are respectively the pole and zero
lines. Their simultaneous satisfaction implies
\begin{equation}
	1-2\Delta=m-\ell\in\mathbb Z,
	\qquad\text{hence}\qquad
	2\Delta\in\mathbb Z.
	\label{eq:S-two-integer-conditions}
\end{equation}
Thus one must distinguish the integer and half-integer classes.

We next show that the half-integer class does not provide two
independent log-free boundary branches. Let
\[
\Delta=\nu+\frac12,
\qquad
\mu=\Delta+m,
\qquad
m\in\mathbb Z_{\ge0},
\]
and introduce \(x=r^{-2}\). The boundary powers are
\(x^{(1-\Delta)/2}\) and \(x^{\Delta/2}\), whose difference is the
integer \(\nu\). The resonant second solution therefore generally
contains a logarithm. After excluding the logarithmic residual piece,
write the source-type branch as
\begin{equation}
	R(x)
	=
	x^{(1-\Delta)/2}(1-x)^{-\mu/2}F(x),
	\qquad
	F(x)=\sum_{s=0}^\infty b_sx^s .
	\label{eq:S-half-source-ansatz}
\end{equation}
The radial equation gives
\begin{equation}
	(1+s)(1+s-\nu)b_{s+1}
	-\frac14
	(-1+m-2s+2\nu)(m-2s+2\nu)b_s=0.
	\label{eq:S-half-recurrence}
\end{equation}
For \(\nu\ge1\), the resonant step \(s=\nu-1\) yields
\[
(m+1)(m+2)b_{\nu-1}=0.
\]
Because \(m\ge0\), one has \(b_{\nu-1}=0\), and successive use of the
recurrence at \(s=\nu-2,\ldots,0\) gives
\begin{equation}
	b_0=b_1=\cdots=b_{\nu-1}=0.
	\label{eq:S-half-zeros}
\end{equation}
Consequently \(F(x)=b_\nu x^\nu+O(x^{\nu+1})\), and hence
\begin{equation}
	R(x)
	\sim
	x^{(1-\Delta)/2+\nu}
	=
	x^{\Delta/2}
	=
	r^{-\Delta}.
\end{equation}
The nominal log-free source branch therefore collapses onto the
response branch:
\begin{equation}
	W[R_{\rm src}^{\rm log-free},R_{\rm resp}]=0.
	\label{eq:S-half-Wzero}
\end{equation}
At the endpoint \(\Delta=1/2\), the two boundary powers coincide
already at leading order, and removing the logarithmic solution again
leaves only one independent branch. \revstart
Thus the half-integer class does not furnish the finite, independent
log-free boundary pair used in this paper. This excludes it from our
two-branch smoothness criterion; a renormalized correlator with
logarithmic boundary data is a separate problem.
\revend

We are left with
\[
\Delta=p\in\mathbb N.
\]
The pole condition then gives
\(\mu=p+m=:N\in\mathbb N\), with \(1\le p\le N\). Therefore the
branch-independent pole-skipping lattice is
\begin{equation}
	\mu=N\in\mathbb N,
	\qquad
	\Delta=p,
	\qquad
	p=1,\ldots,N.
	\label{eq:S-PS-lattice-derived}
\end{equation}
The two outer-horizon indicial roots differ by
\(\mu=N\), so this lattice is precisely the integer Frobenius
resonance sector. At these points \(\Gamma(1-\mu)\) is singular while
\revstart
both $\widehat A$ and $\widehat B$ vanish. This is the formal
common-factor-removed $0/0$; the unregularized boundary pair has
finite nonzero limits along the fixed-mass continuation below.
\revend

\suppsection{Horizon smoothness of the boundary-normalized branches}
\label{sec:S3-boundary-branches}

Introduce
\[
u=\frac{2}{r+1},
\]
so that the AdS boundary and the future horizon correspond respectively
to \(u=0\) and \(u=1\). The boundary-normalized source and response
branches are
\begin{align}
	R_1(r)
	&=
	\chi_\mu(r)(r+1)^{\Delta-1}
	{}_2F_1\left(
	1-\Delta,1-\Delta-\mu;2-2\Delta;u
	\right),
	\label{eq:S-R1-def}
	\\
	R_2(r)
	&=
	\chi_\mu(r)(r+1)^{-\Delta}
	{}_2F_1\left(
	\Delta,\Delta-\mu;2\Delta;u
	\right).
	\label{eq:S-R2-def}
\end{align}
\revstart
At integer \(\Delta\), \(R_1\) means the continuation of this expression
from generic \(\Delta\) at fixed \(\mu\). To define that limit without
a singular hypergeometric denominator, set \(x=r^{-2}\). Direct
substitution into the radial equation, or the quadratic transformation,
gives the equivalent boundary Frobenius basis
\begin{align}
R_1={}&2^{-\mu/2}r^{\Delta-1}(1-r^{-2})^{-\mu/2}
{}_2F_1\!\left(\frac{1-\Delta-\mu}{2},\frac{2-\Delta-\mu}{2};
\frac32-\Delta;r^{-2}\right),
\tag{S26a}\label{eq:S-R1-even}\\
R_2={}&2^{-\mu/2}r^{-\Delta}(1-r^{-2})^{-\mu/2}
{}_2F_1\!\left(\frac{\Delta-\mu}{2},\frac{\Delta+1-\mu}{2};
\Delta+\frac12;r^{-2}\right).
\tag{S26b}\label{eq:S-R2-even}
\end{align}
The source denominator \(3/2-\Delta\) is nonsingular at every integer
\(\Delta\). Its Frobenius powers are \(r^{\Delta-1-2j}\); at
\(\Delta=p\), none equals the response power \(r^{-p}\).
These formulas fix a pure source and pure response basis. In contrast,
terminating the numerator of Eq.~\eqref{eq:S-R1-def} first need not
produce that pure source. For example, at \(\mu=\Delta=1\),
\[
R_1^{\rm cont}=\frac{r}{\sqrt2\sqrt{r^2-1}},\qquad
R_2=\frac{1}{\sqrt2\sqrt{r^2-1}},\qquad
R_1^{\rm term}=\chi_1=R_1^{\rm cont}+R_2.
\]
Since the generic hypergeometric functions approach one as \(u\to0\),
\revend

\[
\frac{R_1}{\chi_\mu}\sim r^{\Delta-1},
\qquad
\frac{R_2}{\chi_\mu}\sim r^{-\Delta},
\]
so \(R_1\) and \(R_2\) are respectively source- and
response-normalized.

\revstart
For both \(u\)-hypergeometric functions, \(c-a-b=\mu\). At
\(\mu=N\in\mathbb N\), the only nonanalytic local term is
\((1-u)^N\log(1-u)\) times an analytic function normalized to one
at \(u=1\). Let \(\mathcal O_j\) be its coefficient in \(R_j/\chi_N\).
The integer-difference connection formula gives
\begin{align}
\mathcal O_2(N,\Delta)&=
\frac{(-1)^{N+1}2^{-\Delta}\Gamma(2\Delta)}
{N!\Gamma(\Delta)\Gamma(\Delta-N)},\\
\mathcal O_1(N,\Delta)&=
\frac{(-1)^{N+1}2^{\Delta-1}\Gamma(2-2\Delta)}
{N!\Gamma(1-\Delta)\Gamma(1-\Delta-N)}.
\end{align}
For the source coefficient the quotient of gamma functions is
continued before specializing \(\Delta\). In particular,
\[
\frac{\Gamma(2-2\Delta)}{\Gamma(1-\Delta)}
=\frac{2^{1-2\Delta}}{\sqrt\pi}\Gamma(3/2-\Delta),
\]
which is finite and nonzero at integer \(\Delta=p\). Thus
\begin{equation}
\mathcal O_2=\frac{(-1)^{N+1}(2\Delta-1)}{2N!}\widehat A,
\qquad
\mathcal O_1=\frac{(-1)^{N+1}(1-2\Delta)}{2N!}\widehat B.
\tag{S29a}\label{eq:S-exact-log-obstructions}
\end{equation}

\revend
Their vanishing is precisely equivalent to the zeros of the
common-factor-removed boundary coefficients introduced in
Sec.~\ref{sec:S2-boundary-coefficients}. Therefore
\begin{equation}
	\boxed{
		\widehat A=0
		\quad\Longleftrightarrow\quad
		\frac{R_2}{\chi_\mu}\in C^\infty,
		\qquad
		\widehat B=0
		\quad\Longleftrightarrow\quad
		\frac{R_1}{\chi_\mu}\in C^\infty
	}
	\label{eq:S-main-dictionary}
\end{equation}
\revstart
for $\mu=N\in\mathbb N$, $\Delta>1/2$, and finite boundary
branches outside $\Delta=1/2,3/2,\ldots$. A source numerator zero
cannot be evaluated independently of its singular normalization.
\revend
 The apparent
interchange follows from boundary normalization: vanishing of the
source coefficient leaves the response-normalized branch, and vice
versa.

At the branch-independent pole-skipping lattice
\[
\mu=N\in\mathbb N,
\qquad
\Delta=p,
\qquad
1\le p\le N,
\]
the residual branches become
\revstart
\begin{align}
	\frac{R_1}{\chi_N}
	&=\frac{r^{p+N-1}}{(r+1)^N}
	{}_2F_1\!\left(\frac{1-p-N}{2},\frac{2-p-N}{2};
	\frac32-p;r^{-2}\right),
	\label{eq:S-R1-polynomial}
	\\
	\frac{R_2}{\chi_N}
	&=
	\frac{r^{N-p}}{(r+1)^N}
	{}_2F_1\!\left(
	\frac{p-N}{2},\frac{p+1-N}{2};p+\frac12;r^{-2}
	\right).
	\label{eq:S-R2-polynomial}
\end{align}
\revend

\revstart
In each of Eqs.~\eqref{eq:S-R1-polynomial} and
\eqref{eq:S-R2-polynomial}, one numerator parameter is a
nonpositive integer, while the denominator parameter is nonsingular
for integer \(p\ge1\). Thus both hypergeometric series terminate
in \(r^{-2}\). Their prefactors are smooth at \(r=1\), including
\(p=N\). Consequently,
\revend

\begin{equation}
	\boxed{
		\frac{R_1}{\chi_N}\in C^\infty,
		\qquad
		\frac{R_2}{\chi_N}\in C^\infty
	}
	\label{eq:S-two-branches-smooth}
\end{equation}
at every pole-skipping point.

\revstart
The branches remain independent. With the convention
\(W[R_1,R_2]=R_1R_2'-R_1'R_2\), their leading boundary coefficients
and Abel's identity give
\begin{equation}
W[R_1,R_2]=\frac{(1-2\Delta)2^{-\mu}}{r^2-1},
\tag{S33a}\label{eq:S-JT-Wronskian}
\end{equation}
where the finite pure-source basis is understood. This is nonzero at
every integer lattice point. Thus the common-factor-removed boundary
condition \(\widehat B/\widehat A=0/0\) is the point at which the
\revend

source- and response-normalized boundary branches simultaneously enter
the horizon residual-smooth subspace.

\suppsection{Boundary path-dependence near the pole-skipping lattice}
\label{sec:S4-boundary-path-dependence}

We now show explicitly how the boundary Green function becomes
path-dependent near the pole-skipping lattice.  Let
\begin{equation}
	\mu=N+\epsilon,
	\qquad
	\Delta=p+\delta,
	\qquad
	N\in\mathbb N,
	\qquad
	p=1,\ldots,N ,
	\label{eq:S-muDelta-epsdelta}
\end{equation}
where $\epsilon$ and $\delta$ are small.  Since the common factor
$2^{-\mu/2}\Gamma(1-\mu)$ cancels in the ratio, we have
\begin{equation}
	\frac{B}{A}
	=
	\frac{\widehat{B}}{\widehat{A}} .
\end{equation}
Using Eq.~\eqref{eq:S-AhatBhat}, this ratio is
\begin{equation}
	\frac{B}{A}
	=
	2^{2\Delta-1}
	\frac{
		\Gamma(1-2\Delta)\Gamma(\Delta)\Gamma(\Delta-\mu)
	}{
		\Gamma(1-\Delta)\Gamma(1-\mu-\Delta)\Gamma(2\Delta-1)
	}.
	\label{eq:S-BA-exact-ratio}
\end{equation}

We use the standard expansion
\begin{equation}
	\Gamma(-q+\eta)
	=
	\frac{(-1)^q}{q! \;\eta}+O(1),
	\qquad
	q=0,1,2,\ldots .
	\label{eq:S-gamma-pole-expansion}
\end{equation}
Near $\Delta=p+\delta$, the ratio of the two boundary-resonant gamma
functions is
\begin{equation}
	\frac{\Gamma(1-2\Delta)}{\Gamma(1-\Delta)}
	=
	(-1)^p
	\frac{(p-1)!}{2(2p-1)!}
	+O(\delta).
	\label{eq:S-boundary-gamma-ratio}
\end{equation}
Also,
\begin{equation}
	\frac{\Gamma(\Delta)}{\Gamma(2\Delta-1)}
	=
	\frac{(p-1)!}{(2p-2)!}
	+O(\delta).
	\label{eq:S-finite-gamma-ratio}
\end{equation}

The singular path-dependence comes from the remaining ratio
\begin{equation}
	\frac{\Gamma(\Delta-\mu)}
	{\Gamma(1-\mu-\Delta)} .
\end{equation}
Indeed,
\begin{equation}
	\Delta-\mu
	=
	p-N+\delta-\epsilon
	=
	-(N-p)+(\delta-\epsilon),
\end{equation}
and
\begin{equation}
	1-\mu-\Delta
	=
	1-N-p-\epsilon-\delta
	=
	-(N+p-1)-(\epsilon+\delta).
\end{equation}
Therefore Eq.~\eqref{eq:S-gamma-pole-expansion} gives
\begin{equation}
	\Gamma(\Delta-\mu)
	=
	\frac{(-1)^{N-p}}
	{(N-p)!(\delta-\epsilon)}
	+O(1),
	\label{eq:S-gamma-Delta-minus-mu}
\end{equation}
and
\begin{equation}
	\Gamma(1-\mu-\Delta)
	=
	\frac{(-1)^{N+p}}
	{(N+p-1)!(\epsilon+\delta)}
	+O(1).
	\label{eq:S-gamma-one-minus-mu-minus-Delta}
\end{equation}
Hence
\begin{equation}
	\frac{\Gamma(\Delta-\mu)}
	{\Gamma(1-\mu-\Delta)}
	=
	\frac{(N+p-1)!}{(N-p)!}
	\frac{\epsilon+\delta}{\delta-\epsilon}
	\left[1+O(\epsilon,\delta)\right].
	\label{eq:S-singular-gamma-ratio}
\end{equation}

Combining Eqs.~\eqref{eq:S-boundary-gamma-ratio},
\eqref{eq:S-finite-gamma-ratio}, and
\eqref{eq:S-singular-gamma-ratio}, we find
\begin{equation}
	\frac{B}{A}
	=
	C_{N,p}
	\frac{\epsilon+\delta}{\delta-\epsilon}
	\left[1+O(\epsilon,\delta)\right],
	\label{eq:S-BA-path-dependent}
\end{equation}
where
\begin{equation}
	C_{N,p}
	=
	(-1)^p
	2^{2p-2}
	\frac{
		[(p-1)!]^2 (N+p-1)!
	}{
		(2p-1)!(2p-2)!(N-p)!
	}.
	\label{eq:S-CNp}
\end{equation}
The constant $C_{N,p}$ is finite and nonzero for $p=1,\ldots,N$.

Equation~\eqref{eq:S-BA-path-dependent} shows that the boundary Green
function has no unique value at the pole-skipping point.  For example,
\begin{equation}
	\lim_{\epsilon\to0}\lim_{\delta\to0}
	\frac{B}{A}
	=
	-C_{N,p},
	\qquad
	\lim_{\delta\to0}\lim_{\epsilon\to0}
	\frac{B}{A}
	=
	C_{N,p}.
	\label{eq:S-noncommuting-limits}
\end{equation}
Thus
\begin{equation}
	\lim_{\epsilon\to0}\lim_{\delta\to0}
	\frac{B}{A}
	\neq
	\lim_{\delta\to0}\lim_{\epsilon\to0}
	\frac{B}{A}.
\end{equation}
More generally, along a straight path
\begin{equation}
	\delta=\lambda\epsilon,
	\qquad
	\lambda\neq1,
\end{equation}
one obtains
\begin{equation}
	\frac{B}{A}
	\longrightarrow
	C_{N,p}\frac{1+\lambda}{\lambda-1}.
	\label{eq:S-straight-path-limit}
\end{equation}
Therefore different approaches to the same point
$(\mu,\Delta)=(N,p)$ give different boundary values.  This is the
boundary manifestation of the pole-skipping ambiguity.  In the next
section we show how the same path-dependence appears in the bulk as the
unfixed resonant Frobenius coefficient.

\suppsection{Frobenius recurrence, compatibility, and the \(a_N\) freedom}
\label{sec:S5-frobenius-recurrence}

At the integer horizon resonance \(\mu=N\in\mathbb N\), write
\[
R(r)=\chi_N(r)F(z),
\qquad
z=\frac{1-r}{2}.
\]
The residual equation is
\begin{equation}
	z(1-z)F''+(1-N-2z)F'
	+\Delta(\Delta-1)F=0.
	\label{eq:S5-residual-equation-general}
\end{equation}
For the \(\lambda=0\) Frobenius branch
\begin{equation}
	F(z)=\sum_{k=0}^{\infty}a_kz^k,
	\qquad
	a_0\neq0,
	\label{eq:S5-F-series}
\end{equation}
Eq.~\eqref{eq:S5-residual-equation-general} gives
\begin{equation}
	k(k-N)a_k
	=
	(k-1+\Delta)(k-\Delta)a_{k-1},
	\qquad k\ge1.
	\label{eq:S5-recurrence-general}
\end{equation}
For \(0\le k\le N-1\), this yields
\begin{equation}
	a_k
	=
	a_0
	\frac{(\Delta)_k(1-\Delta)_k}
	{k!(1-N)_k}.
	\label{eq:S5-ak-formal-before-resonance}
\end{equation}

At the resonant step \(k=N\), the recurrence reduces to
\begin{equation}
	0=(N-1+\Delta)(N-\Delta)a_{N-1}.
	\label{eq:S5-compatibility-step}
\end{equation}
For \(\Delta>0\), the compatibility condition is therefore
\((N-\Delta)a_{N-1}=0\). If \(\Delta\neq N\), Eq.~\eqref{eq:S5-ak-formal-before-resonance}
shows that compatibility requires
\[
(1-\Delta)_{N-1}=0,
\]
and hence \(\Delta=1,\ldots,N-1\). Together with the separate
possibility \(\Delta=N\), the log-free compatibility values are
\begin{equation}
	\Delta=p,
	\qquad
	p=1,\ldots,N.
	\label{eq:S5-Delta-p-derived}
\end{equation}

We first consider \(1\le p<N\). The recurrence becomes
\begin{equation}
	k(k-N)a_k
	=
	(k-1+p)(k-p)a_{k-1}.
	\label{eq:S5-recurrence-p}
\end{equation}
For \(0\le k\le p-1\),
\begin{equation}
	a_k
	=
	a_0
	\frac{(p)_k(1-p)_k}
	{k!(1-N)_k}.
	\label{eq:S5-first-region}
\end{equation}
At \(k=p\), the factor \(k-p\) vanishes, so \(a_p=0\);
successive use of the recurrence gives
\begin{equation}
	a_p=a_{p+1}=\cdots=a_{N-1}=0.
	\label{eq:S5-middle-zero-region}
\end{equation}
The resonant equation at \(k=N\) is then identically \(0=0\), leaving
\(a_N\) as a new independent Frobenius datum. The higher coefficients
generated by \(a_N\) form the second local branch, and the general
residual solution is
\begin{equation}
	F(z)=a_0G_1(z)+a_N z^N G_2(z),
	\qquad 1\le p<N,
	\label{eq:S5-general-solution-p-less-N}
\end{equation}
where
\begin{align}
	G_1(z)
	&=
	{}_2F_1(p,1-p;1-N;z)
	=
	\sum_{k=0}^{p-1}
	\frac{(p)_k(1-p)_k}
	{k!(1-N)_k}z^k,
	\label{eq:S5-G1}
	\\
	G_2(z)
	&=
	{}_2F_1(N+p,N+1-p;N+1;z).
	\label{eq:S5-G2}
\end{align}

The physical meaning of \(a_N\) follows after restoring the full radial
field. Near \(r=1\),
\[
\chi_N(r)\sim(r-1)^{-N/2},
\qquad
z=-\frac{r-1}{2}.
\]
Since \(G_1(0)=G_2(0)=1\),
\begin{equation}
	\chi_N G_1
	\sim(r-1)^{-N/2},
	\qquad
	\chi_N z^N G_2
	\sim(r-1)^{+N/2},
	\label{eq:S5-radial-branches}
\end{equation}
\revstart
up to nonzero constant factors. The second term has outgoing radial
behavior, but at resonance both residuals are analytic. The selection
must be tied to a specified limit. Hold \(\Delta=p\) fixed in
Eq.~\eqref{eq:S-ingoing-solution}: its nonresonant residual terminates
at degree \(p-1\). Since all denominators for \(k\le p-1<N\)
stay nonzero as \(\mu\to N\), the limit is \(G_1\) and hence
\revend

\begin{equation}
	a_N=0,
	\qquad
	1\le p<N.
	\label{eq:S5-aN-zero}
\end{equation}
\revstart
Thus residual $C^\infty$ regularity leaves an additional resonant
datum, while the specified fixed-$\Delta$ continuation sets it to zero.
\revend

The endpoint \(p=N\) is treated separately. In this case
\begin{equation}
	k(k-N)a_k
	=
	(k-1+N)(k-N)a_{k-1}.
	\label{eq:S5-recurrence-endpoint}
\end{equation}
\revstart
For \(k\ne N\), cancellation of \(k-N\) gives
\(ka_k=(k-1+N)a_{k-1}\). At \(k=N\), however, the original
equation is \(0=0\): \(a_N\) is free and the cancelled relation
must not be imposed across this step. At fixed \(\Delta=N\), the
nonresonant series terminates at \(k=N-1\), so its actual limit is
\begin{equation}
F_{N,N}^{\rm can}(z)=a_0T_{N-1}(z),\qquad
T_{N-1}(z)=\sum_{k=0}^{N-1}\frac{(N)_k}{k!}z^k.
\label{eq:S5-endpoint-branch}
\end{equation}
This gives \(a_N=0\). The full endpoint solution space is
\begin{equation}
F(z)=a_0T_{N-1}(z)+a_Nz^N{}_2F_1(2N,1;N+1;z).
\tag{S67a}\label{eq:S5-endpoint-general}
\end{equation}
The response-normalized branch instead satisfies
\(2^NR_2/\chi_N=(1-z)^{-N}\). Splitting its binomial series gives
the exact identity
\begin{equation}
2^NR_2-R_{N,N}^{\rm can}
=\frac{(N)_N}{N!}\chi_Nz^N{}_2F_1(2N,1;N+1;z).
\tag{S67b}\label{eq:S5-endpoint-tail}
\end{equation}
Thus the endpoint response and canonical modes differ by a nonzero
second resonant direction. For \(N=1\), \(R_{1,1}^{\rm can}=\chi_1\)
and \(R_2=\chi_1/(r+1)\); they are not constant multiples.

\revend
\suppsection{Ordered auxiliary limit for the polynomial branch}
\label{sec:S6-ordered-auxiliary-limit}

\revstart
Throughout this section $1\le p\le N$. The endpoint uses the
finite polynomial established in Sec.~\ref{sec:S5-frobenius-recurrence};
the same coefficient formulas apply there.
\revend
  The resonant \(a_0\)-branch is
\begin{equation}
	G_1(z)={}_2F_1(p,1-p;1-N;z).
	\label{eq:S6-G1-polynomial}
\end{equation}
Because its parameters differ by integers, its boundary coefficients
cannot be obtained by direct substitution into the generic
hypergeometric connection formula.  We therefore introduce
\begin{equation}
	F_{\alpha,\beta}(z)
	=
	{}_2F_1(p,1-p+\alpha;1-N+\beta;z)
	\label{eq:S6-aux-family}
\end{equation}
and recover the same terminating branch through the ordered limit
\begin{equation}
	G_1(z)
	=
	\lim_{\beta\to0}\lim_{\alpha\to0}
	F_{\alpha,\beta}(z).
	\label{eq:S6-ordered-limit}
\end{equation}
The numerator regulator must be removed first: at fixed nonzero
\(\beta\), the limit \(\alpha\to0\) restores the terminating parameter
\(1-p\), after which \(\beta\to0\) restores the resonant denominator
parameter.

For generic \(\alpha,\beta\), the coefficient of the source-side
asymptotic power \((-z)^{p-1-\alpha}\) is
\begin{equation}
	C_b(\alpha,\beta)
	=
	\frac{
		\Gamma(1-N+\beta)\Gamma(2p-1-\alpha)
	}{
		\Gamma(p)\Gamma(p-N+\beta-\alpha)
	}.
	\label{eq:S6-Cb}
\end{equation}
Using the gamma-function expansion near negative integers gives
\begin{equation}
	C_b(\alpha,\beta)
	=
	C_b^{\rm can}
	\frac{\beta-\alpha}{\beta}
	\bigl[1+O(\alpha,\beta)\bigr],
	\qquad
	C_b^{\rm can}
	=
	(-1)^{p-1}
	\frac{\Gamma(2p-1)}{\Gamma(p)}
	\frac{(N-p)!}{(N-1)!}.
	\label{eq:S6-Cb-near}
\end{equation}
Consequently,
\begin{equation}
	\lim_{\beta\to0}\lim_{\alpha\to0}
	C_b(\alpha,\beta)
	=
	C_b^{\rm can},
\end{equation}
whereas, for example, the path \(\alpha=\beta\) gives zero at leading
order.  The auxiliary limit is therefore path dependent unless the
terminating branch is restored first.

Since \(-z=(r-1)/2\) and
\(\chi_N(r)\sim2^{-N/2}\) as \(r\to\infty\), the source coefficient of
\(R_0(r)=\chi_N(r)G_1(z)\) is
\begin{equation}
	A_0
	=
	(-1)^{p-1}
	2^{1-p-N/2}
	\frac{(2p-2)!(N-p)!}
	{(p-1)!(N-1)!}.
	\label{eq:S6-A0}
\end{equation}

\revstart
The auxiliary family restores the polynomial but is not itself a
fixed-mass JT family while its regulators are nonzero. Boundary data
are measured in the continued \(r^{-2}\) Frobenius basis of
Sec.~\ref{sec:S3-boundary-branches}. At integer \(p\), the source
powers \(r^{p-1-2j}\) never equal \(r^{-p}\), so the two coefficients
are separately identifiable. The complete boundary pair of the
canonical polynomial is equivalently obtained from the fixed-\(p\)
ingoing family:
\revend

\begin{equation}
	A_0=\lim_{\mu\to N}A(\mu,p),
	\qquad
	B_0=\lim_{\mu\to N}B(\mu,p).
	\label{eq:S6-A0B0-canonical-limit}
\end{equation}
The first equality in Eq.~\eqref{eq:S6-A0B0-canonical-limit}
follows directly from Eq.~\eqref{eq:S-A-coeff} and reproduces the
independently obtained result in Eq.~\eqref{eq:S6-A0}.  It remains to
evaluate the response coefficient \(B_0\).  At fixed integer \(p\),
the apparently singular quotient entering \(B(\mu,p)\) is evaluated
through its meromorphic extension.  The Gamma-function duplication
identity gives
\begin{equation}
	\frac{\Gamma(1-2\Delta)}
	{\Gamma(1-\Delta)}
	=
	\frac{2^{-2\Delta}}{\sqrt{\pi}}
	\Gamma\left(\frac12-\Delta\right).
	\label{eq:S6-gamma-duplication-ratio}
\end{equation}
The right-hand side is finite at \(\Delta=p\), and hence
\begin{equation}
	\left.
	\frac{\Gamma(1-2\Delta)}
	{\Gamma(1-\Delta)}
	\right|_{\Delta=p}^{\rm mer}
	=
	\frac{2^{-2p}}{\sqrt{\pi}}
	\Gamma\left(\frac12-p\right)
	=
	(-1)^p\frac{(p-1)!}{2(2p-1)!}.
	\label{eq:S6-gamma-ratio-fixed-p}
\end{equation}
Thus \(p\) remains fixed throughout the physical continuation, and
the only parameter limit is \(\mu\to N\).  In particular,
\begin{equation}
	B(\mu,p)
	=
	(-1)^p
	2^{p-\mu/2-1}
	\frac{(p-1)!}{(2p-1)!}
	\frac{\Gamma(1-\mu)}
	{\Gamma(1-\mu-p)}.
	\label{eq:S6-B-fixed-p}
\end{equation}
Taking \(\mu\to N\) in Eq.~\eqref{eq:S6-B-fixed-p} and using
\begin{equation}
	\lim_{\mu\to N}
	\frac{\Gamma(1-\mu)}{\Gamma(1-\mu-p)}
	=
	(-1)^p\frac{(N+p-1)!}{(N-1)!}
\end{equation}
gives
\begin{equation}
	B_0
	=
	2^{p-N/2-1}
	\frac{(p-1)!(N+p-1)!}
	{(2p-1)!(N-1)!}.
	\label{eq:S6-B0}
\end{equation}
Thus
\begin{equation}
	R_0(r)=\chi_N(r)G_1(z)
	\sim
	A_0r^{p-1}+B_0r^{-p},
	\label{eq:S6-R0-boundary}
\end{equation}
and the ordered auxiliary limit selects the same resonant branch as continuation of the nonresonant ingoing solution.

\suppsection{Why continuation of the nonresonant ingoing solution gives \(a_N=0\)}
\label{sec:S7-why-aN-zero}

Throughout this section we assume
\revstart
\[
\mu=N\in\mathbb N,\quad\Delta=p,\quad1\le p\le N.
\]
The endpoint polynomial is the corrected expression in
Sec.~\ref{sec:S5-frobenius-recurrence}. At resonance, the residual
\revend

solution is
\begin{equation}
	F(z)=a_0G_1(z)+a_N z^N G_2(z),
	\qquad
	G_2(z)={}_2F_1(N+p,N+1-p;N+1;z).
	\label{eq:S7-residual-general}
\end{equation}
As shown in Sec.~\ref{sec:S5-frobenius-recurrence}, restoring the full
radial field gives
\begin{equation}
	\chi_N G_1\sim(r-1)^{-N/2},
	\qquad
	\chi_N z^N G_2\sim(r-1)^{+N/2}.
	\label{eq:S7-radial-branches}
\end{equation}
\revstart
The \(a_N\)-term has outgoing radial behavior. This observation does
not determine a parameter-space path at exact resonance. The
fixed-\(\Delta=p\) finite-series argument in Sec.~\ref{sec:S5-frobenius-recurrence}
selects \(a_N=0\); we now check the same specified limit using its
boundary data.
\revend

Define
\[
R_0(r)=\chi_N(r)G_1(z),
\qquad
R_N(r)=\chi_N(r)z^N G_2(z).
\]
Their boundary expansions are
\begin{equation}
	R_0\sim A_0r^{p-1}+B_0r^{-p},
	\qquad
	R_N\sim A_Nr^{p-1}+B_Nr^{-p},
	\label{eq:S7-resonant-boundary-expansions}
\end{equation}
where \(A_0\) and \(B_0\) are given in
Eqs.~\eqref{eq:S6-A0} and \eqref{eq:S6-B0}, while
\begin{align}
	A_N
	&=
	(-1)^N
	2^{1-p-N/2}
	\frac{N!(2p-2)!}
	{(p-1)!(N+p-1)!},
	\label{eq:S7-AN}
	\\
	B_N
	&=
	(-1)^{N-p}
	2^{p-N/2-1}
	\frac{N!(p-1)!}
	{(2p-1)!(N-p)!}.
	\label{eq:S7-BN}
\end{align}
\revstart
Both coefficients are finite and nonzero for $1\le p\le N$.
\revend

By linearity, after fixing the horizon normalization \(a_0=1\), the
general resonant solution has boundary data
\begin{equation}
	(A_{\rm res},B_{\rm res})
	=
	(A_0,B_0)+a_N(A_N,B_N).
	\label{eq:S7-linear-boundary-data}
\end{equation}
On the other hand, continuation of the nonresonant ingoing solution derived in Sec.~\ref{sec:S6-ordered-auxiliary-limit}
satisfies
\begin{equation}
	\lim_{\mu\to N}
	(A(\mu,p),B(\mu,p))
	=
	(A_0,B_0).
	\label{eq:S7-nonres-to-A0B0}
\end{equation}
Matching the resonant solution to this continuation limit therefore
requires
\[
a_N(A_N,B_N)=(0,0).
\]
Since \((A_N,B_N)\neq(0,0)\), one obtains
\begin{equation}
	a_N=0.
	\label{eq:S7-aN-zero-matching}
\end{equation}

\revstart
The fixed-$\Delta$ finite-series limit and its boundary-data check
therefore select the same resonant solution:
\revend

\begin{equation}
	F_{\rm can}(z)=G_1(z),
	\qquad
	a_N=0.
	\label{eq:S7-canonical-representative}
\end{equation}

\revstart

The physical motivation for this prescription is to vary frequency
within one scalar theory: \(m^2=p(p-1)\) and the chosen
standard-quantization branch \(\Delta=p\) remain fixed. It is not a
claim that causality forbids every other parameter-space path.
Indeed, with \(\mu=N+\epsilon\), \(\Delta=p+\sigma\epsilon\), the
\(z^N\) coefficient of the nonresonant ingoing residual has limit
\begin{equation}
a_N\longrightarrow \sigma K_{N,p},\qquad
K_{N,p}=(-1)^{N-p}\frac{(N+p-1)!(N-p)!}{N!(N-1)!}.
\tag{S90a}\label{eq:S7-general-path-datum}
\end{equation}
This follows by expanding the products
\((\Delta)_N(1-\Delta)_N/[N!(1-\mu)_N]\) to first order.
All lower coefficients approach those of \(G_1\), while higher
coefficients are generated from this \(a_N\); the local limit is
\(G_1+\sigma K_{N,p}z^NG_2\). The identities
\(K_{N,p}A_N=-A_0\) and \(K_{N,p}B_N=B_0\) then give
\[
\frac{B}{A}\longrightarrow\frac{B_0}{A_0}\frac{1+\sigma}{1-\sigma}
=C_{N,p}\frac{1+\sigma}{\sigma-1},
\]
for \(\sigma\ne1\), in agreement with Sec.~S4. At \(N=p=1\),
the residual is \(1+\sigma z/(1-z)\): \(\sigma=0\) selects the
canonical mode, while \(\sigma=1\) selects \(2R_2\).
Every nonresonant member used in this example is ingoing, although
\(\Delta\) and hence the mass vary when \(\sigma\ne0\).

\suppsection{Thermal and representation-theoretic interpretation}
\revend

\label{sec:S8-thermal-representation}

\suppsubsection{Euclidean thermal label and Lorentzian resonance}
\label{subsec:S8-euclidean-label}

Consider a static nonextremal black hole
\begin{equation}
	ds^2=-f(r)dt^2+\frac{dr^2}{f(r)},
	\qquad
	f(r_+)=0,
	\qquad
	\kappa_+=\frac12 f'(r_+).
\end{equation}
Near the outer horizon,
\(f(r)\simeq2\kappa_+(r-r_+)\). After the Wick rotation
\(t=-i\tau\) and the change of variable
\(r-r_+=\kappa_+R^2/2\), the Euclidean metric becomes
\begin{equation}
	ds_E^2\simeq dR^2+\kappa_+^2R^2d\tau^2
	=dR^2+R^2d\theta^2,
	\qquad
	\theta=\kappa_+\tau .
	\label{eq:S8-Euclidean-cap}
\end{equation}
Smoothness at \(R=0\) requires
\(\theta\sim\theta+2\pi\), or equivalently
\[
\tau\sim\tau+\beta,
\qquad
\beta=\frac{2\pi}{\kappa_+}.
\]
A smooth bosonic field therefore admits angular harmonics
\(e^{im\theta}=e^{i\omega_E\tau}\), with
\begin{equation}
	\omega_E=m\kappa_+
	=\frac{2\pi m}{\beta},
	\qquad
	m\in\mathbb Z .
	\label{eq:S8-matsubara}
\end{equation}

For the Lorentzian convention
\(\Phi_L=e^{-i\omega t}R_\omega(r)\), the same Wick rotation gives
\(e^{-i\omega t}\to e^{-\omega\tau}\). Comparison with
\(e^{i\omega_E\tau}\) yields the retarded continuation
\begin{equation}
	\omega_E=i\omega .
	\label{eq:S8-omegaE}
\end{equation}
Hence the Euclidean thermal label satisfies
\begin{equation}
	\frac{i\omega}{\kappa_+}=m.
	\label{eq:S8-thermal-label}
\end{equation}

The two Lorentzian near-horizon radial behaviors are
\begin{equation}
	R(r)\sim
	(r-r_+)^{-i\omega/(2\kappa_+)},
	\qquad
	R(r)\sim
	(r-r_+)^{+i\omega/(2\kappa_+)}.
\end{equation}
Their Frobenius indices differ by \(i\omega/\kappa_+\). The integer
resonance condition is therefore
\begin{equation}
	\frac{i\omega}{\kappa_+}=N,
	\qquad
	N\in\mathbb N
	\label{eq:S8-lorentzian-resonance}
\end{equation}
in \revstart
the positive-$\mu$ sector considered here ($\omega=-iN\kappa_+$).
\revend

Consequently,
\begin{equation}
	m=N.
\end{equation}
\revstart
This compares integer labels only. The ingoing power at $N>0$
becomes $R^{-N}$ on the Euclidean cap and is singular there, whereas
a smooth angular harmonic requires $R^{|N|}$. Lorentzian residual
smoothness is therefore not full Euclidean cap regularity.

For JT/AdS$_2$, $f(r)=r^2-1$, $r_+=1$, and
$\kappa_+=1$, so the resonance condition reduces to
\revend

\begin{equation}
	\mu=i\omega=N.
\end{equation}

\suppsubsection{\(SL(2,\mathbb R)\) generators and Casimir}
\label{subsec:S8-sl2-generators}

For the JT/AdS$_2$ black-hole metric
\begin{equation}
	ds^2=-(r^2-1)dt^2+\frac{dr^2}{r^2-1},
\end{equation}
let \(s(r)=\sqrt{r^2-1}\). A convenient basis of Killing generators is
\begin{align}
	J_0&=-\partial_t,
	\nonumber\\
	J_-&=e^t\left(
	s\,\partial_r-\frac{r}{s}\partial_t
	\right),
	\nonumber\\
	J_+&=-e^{-t}\left(
	s\,\partial_r+\frac{r}{s}\partial_t
	\right).
	\label{eq:S8-sl2-generators}
\end{align}
They obey
\begin{equation}
	[J_0,J_\pm]=\pm J_\pm,
	\qquad
	[J_-,J_+]=2J_0.
	\label{eq:S8-sl2-algebra}
\end{equation}

With the lowest-weight convention, the quadratic Casimir is
\begin{equation}
	\mathcal C
	=
	J_0(J_0-1)-J_+J_-
	=
	\partial_r\!\left[(r^2-1)\partial_r\right]
	-\frac{1}{r^2-1}\partial_t^2.
	\label{eq:S8-casimir-box}
\end{equation}
Thus \(\mathcal C\) coincides with the scalar d'Alembertian, and the
Klein--Gordon equation with \(m^2=\Delta(\Delta-1)\) becomes
\begin{equation}
	\mathcal C\Phi
	=
	\Delta(\Delta-1)\Phi.
	\label{eq:S8-casimir-eigenvalue}
\end{equation}

\suppsubsection{Lowest-weight mode and the response ladder}
\label{subsec:S8-lowest-weight}

At the endpoint
\[
\Delta=\mu=n,
\]
\revstart
the response-normalized mode, distinct from the fixed-mass canonical
continuation, is
\revend

\begin{equation}
	\Phi^{\rm resp}_{n,0}(t,r)
	=
	C_n e^{-nt}(r^2-1)^{-n/2}.
	\label{eq:S8-lowest-mode}
\end{equation}
For later use, define
\begin{equation}
	E_N(t,r)
	:=
	e^{-Nt}(r^2-1)^{-N/2}.
	\label{eq:S8-EN-def}
\end{equation}
Then \(\Phi^{\rm resp}_{n,0}=C_nE_n\). Using the generators of
Sec.~\ref{subsec:S8-sl2-generators}, one finds directly
\begin{equation}
	J_0E_N=NE_N,
	\qquad
	J_-E_N=0.
	\label{eq:S8-EN-lowest}
\end{equation}
Indeed, with \(s=\sqrt{r^2-1}\),
\[
J_-E_N
=
e^{-(N-1)t}
\left[
s\,\partial_r s^{-N}
+\frac{Nr}{s}s^{-N}
\right]
=0.
\]
Therefore
\begin{equation}
	J_-\Phi^{\rm resp}_{n,0}=0,
	\qquad
	J_0\Phi^{\rm resp}_{n,0}=n\Phi^{\rm resp}_{n,0},
	\qquad
	\mathcal C\Phi^{\rm resp}_{n,0}
	=
	n(n-1)\Phi^{\rm resp}_{n,0}.
	\label{eq:S8-response-lowest-condition}
\end{equation}
The last relation also follows algebraically from
\(\mathcal C=J_0(J_0-1)-J_+J_-\).

Acting repeatedly with \(J_+\) generates the standard lowest-weight
module associated with the response-normalized polynomial branch.
This module should not be identified with the full two-dimensional
radial solution space: the same auxiliary radial equation also admits
a source companion branch, which is not generated from
\(\Phi^{\rm resp}_{n,0}\) by this lowest-weight construction.

\suppsubsection{The auxiliary polynomial equation and its two polynomial branches}
\label{subsec:S8-two-polynomial-branches}

We first recall the radial differential operator
\begin{equation}
	\mathcal D_{\Delta,\mu}
	:=
	(1-r^2)\frac{d^2}{dr^2}
	-
	2r\frac{d}{dr}
	+
	\left[
	\Delta(\Delta-1)
	-
	\frac{\mu^2}{1-r^2}
	\right].
	\label{eq:S8-radial-operator-ladder}
\end{equation}
The radial equation is
\begin{equation}
	\mathcal D_{\Delta,\mu}R(r)=0.
\end{equation}
In particular,
\begin{equation}
	\mathcal D_{n,n+q}
	=
	(1-r^2)\frac{d^2}{dr^2}
	-
	2r\frac{d}{dr}
	+
	\left[
	n(n-1)
	-
	\frac{(n+q)^2}{1-r^2}
	\right].
	\label{eq:S8-Dnnq-explicit}
\end{equation}
Here \(n=p\) is the fixed conformal weight, while
\(N=n+q\) remains the resonance label used for \(\mu\).
Fix
\begin{equation}
	\Delta=n,
	\qquad
	\mu=N=n+q,
	\qquad
	q=0,1,2,\ldots .
\end{equation}
Let
\begin{equation}
	s(r):=\sqrt{r^2-1}.
\end{equation}
For a radial function of the form
\begin{equation}
	R(r)
	=
	s(r)^{-N}P(r)
	=
	(r^2-1)^{-N/2}P(r),
	\label{eq:S8-radial-polynomial-factorization}
\end{equation}
substitution into the radial equation gives a polynomial equation for
\(P(r)\).  We define
\begin{equation}
	\mathcal L_{n,q}
	:=
	(r^2-1)\frac{d^2}{dr^2}
	-
	2(n+q-1)r\frac{d}{dr}
	+
	q(2n+q-1).
	\label{eq:S8-Lnq-definition}
\end{equation}
Then the exact identity is
\begin{equation}
	\mathcal D_{n,n+q}
	\left[
	s(r)^{-(n+q)}P(r)
	\right]
	=
	-
	s(r)^{-(n+q)}
	\mathcal L_{n,q}P(r).
	\label{eq:S8-D-to-L-identity}
\end{equation}
Therefore
\begin{equation}
	\mathcal D_{n,n+q}
	\left[
	s(r)^{-(n+q)}P(r)
	\right]
	=0
\end{equation}
is equivalent to
\begin{equation}
	\mathcal L_{n,q}P(r)=0.
	\label{eq:S8-auxiliary-equation}
\end{equation} 
The equation \eqref{eq:S8-auxiliary-equation} is a Gegenbauer equation.
The standard Gegenbauer equation is
\begin{equation}
	(1-r^2)y''
	-
	(2\alpha+1)r y'
	+
	m(m+2\alpha)y
	=
	0.
	\label{eq:S8-Gegenbauer-equation}
\end{equation}
Multiplying by \(-1\), and setting
\begin{equation}
	\alpha
	=
	\frac12-n-q,
	\label{eq:S8-alpha-definition}
\end{equation}
we get
\begin{equation}
	(r^2-1)y''
	-
	2(n+q-1)r y'
	-
	m(m+2\alpha)y
	=
	0.
\end{equation}
Thus the coefficient of \(y\) agrees with that in
\eqref{eq:S8-auxiliary-equation} if
\begin{equation}
	-m(m+2\alpha)
	=
	q(2n+q-1).
\end{equation}
Equivalently,
\begin{equation}
	\bigl(m-q\bigr)
	\bigl(m-(2n+q-1)\bigr)
	=
	0 .
\end{equation}
Hence there are two polynomial degrees:
\begin{equation}
	m_{\rm resp}=q,
	\qquad
	m_{\rm src}=2n+q-1.
	\label{eq:S8-two-degrees}
\end{equation}

We use monic Gegenbauer polynomials
\begin{equation}
	\mathcal G_m^{(\alpha)}(r)
	:=
	\frac{m!}{2^m(\alpha)_m}
	C_m^{(\alpha)}(r),
	\qquad
	\mathcal G_m^{(\alpha)}(r)
	=
	r^m+\cdots .
	\label{eq:S8-monic-Gegenbauer}
\end{equation}
Their finite series form is
\begin{equation}
	\mathcal G_m^{(\alpha)}(r)
	=
	\frac{m!}{2^m(\alpha)_m}
	\sum_{j=0}^{\lfloor m/2\rfloor}
	(-1)^j
	\frac{\Gamma(m-j+\alpha)}
	{\Gamma(\alpha) j! (m-2j)!}
	(2r)^{m-2j}.
	\label{eq:S8-monic-Gegenbauer-series}
\end{equation}
Here \(\lfloor m/2\rfloor\) is the largest integer not exceeding
\(m/2\).  It is the upper limit of the summation index \(j\), not the
degree of the polynomial.  The degree is \(m\).

We define the two monic basis polynomials by
\begin{align}
	P_{\rm resp}^{(n,q)}(r)
	&:=
	\mathcal G_q^{\left(\frac12-n-q\right)}(r),
	\label{eq:S8-Presp-definition}
	\\
	P_{\rm src}^{(n,q)}(r)
	&:=
	\mathcal G_{2n+q-1}^{\left(\frac12-n-q\right)}(r).
	\label{eq:S8-Psrc-definition}
\end{align}
Equivalently, their explicit finite series are
\begin{align}
	P_{\rm resp}^{(n,q)}(r)
	&=
	\frac{q!}
	{2^q\left(\frac12-n-q\right)_q}
	\sum_{j=0}^{\lfloor q/2\rfloor}
	(-1)^j
	\frac{
		\Gamma\left(q-j+\frac12-n-q\right)
	}
	{
		\Gamma\left(\frac12-n-q\right)
		j!(q-2j)!
	}
	(2r)^{q-2j},
	\label{eq:S8-Presp-series}
	\\
	P_{\rm src}^{(n,q)}(r)
	&=
	\frac{(2n+q-1)!}
	{2^{2n+q-1}
		\left(\frac12-n-q\right)_{2n+q-1}}
	\sum_{j=0}^{\lfloor(2n+q-1)/2\rfloor}
	(-1)^j
	\frac{
		\Gamma\left(2n+q-1-j+\frac12-n-q\right)
	}
	{
		\Gamma\left(\frac12-n-q\right)
		j!(2n+q-1-2j)!
	}
	(2r)^{2n+q-1-2j}.
	\label{eq:S8-Psrc-series}
\end{align}

Both satisfy the same auxiliary equation:
\begin{equation}
	\mathcal L_{n,q}P_{\rm resp}^{(n,q)}=0,
	\qquad
	\mathcal L_{n,q}P_{\rm src}^{(n,q)}=0.
	\label{eq:S8-two-basis-solve-L}
\end{equation}
Therefore the corresponding full radial functions
\begin{align}
	R_{\rm resp}^{(n,q)}(r)
	&:=
	s(r)^{-(n+q)}P_{\rm resp}^{(n,q)}(r),
	\label{eq:S8-Rresp-definition}
	\\
	R_{\rm src}^{(n,q)}(r)
	&:=
	s(r)^{-(n+q)}P_{\rm src}^{(n,q)}(r)
	\label{eq:S8-Rsrc-definition}
\end{align}
satisfy
\begin{equation}
	\mathcal D_{n,n+q}R_{\rm resp}^{(n,q)}=0,
	\qquad
	\mathcal D_{n,n+q}R_{\rm src}^{(n,q)}=0.
	\label{eq:S8-two-radial-solutions}
\end{equation}

Their boundary behaviors are different.  Since
\begin{equation}
	P_{\rm resp}^{(n,q)}(r)\sim r^q,
	\qquad
	P_{\rm src}^{(n,q)}(r)\sim r^{2n+q-1},
\end{equation}
we have
\begin{align}
	R_{\rm resp}^{(n,q)}(r)
	&\sim
	r^{-(n+q)}r^q
	=
	r^{-n},
	\label{eq:S8-Rresp-boundary}
	\\
	R_{\rm src}^{(n,q)}(r)
	&\sim
	r^{-(n+q)}r^{2n+q-1}
	=
	r^{n-1}.
	\label{eq:S8-Rsrc-boundary}
\end{align}
Thus \(R_{\rm resp}^{(n,q)}\) is the response branch and
\(R_{\rm src}^{(n,q)}\) is the source branch for \(\Delta=n\).

\suppsubsection{Ladder action on the two polynomial branches}
\label{subsec:S8-ladder-action-two-branches}

For \(N=n+q\), the Killing generators act on
\(E_NP(r)\) as
\begin{align}
	J_+\bigl(E_NP\bigr)
	&=
	E_{N+1}\mathcal A_N^+P,
	&
	\mathcal A_N^+
	&=
	2Nr-(r^2-1)\frac{d}{dr},
	\label{eq:S8-Jplus-ANplus}
	\\
	J_-\bigl(E_NP\bigr)
	&=
	E_{N-1}\mathcal A_N^-P,
	&
	\mathcal A_N^-
	&=
	\frac{d}{dr}.
	\label{eq:S8-Jminus-ANminus}
\end{align}
A direct differentiation gives the intertwining identities
\begin{equation}
	\mathcal L_{n,q+1}\mathcal A_{n+q}^+
	=
	\mathcal A_{n+q}^+\mathcal L_{n,q},
	\qquad
	\mathcal L_{n,q-1}\mathcal A_{n+q}^-
	=
	\mathcal A_{n+q}^-\mathcal L_{n,q}.
	\label{eq:S8-intertwining-identities}
\end{equation}
Hence \(\mathcal A_{n+q}^+\) and
\(\mathcal A_{n+q}^-\) map polynomial solutions at level \(q\)
to polynomial solutions at levels \(q+1\) and \(q-1\), respectively.

Recall
\[
P_{\rm resp}^{(n,q)}
=
\mathcal G_q^{(\alpha_q)},
\qquad
P_{\rm src}^{(n,q)}
=
\mathcal G_{2n+q-1}^{(\alpha_q)},
\qquad
\alpha_q=\frac12-n-q,
\]
where both polynomials are monic.

For the response branch,
\(P_{\rm resp}^{(n,q)}=r^q+\cdots\), and therefore
\[
\mathcal A_{n+q}^+P_{\rm resp}^{(n,q)}
=
(2n+q)r^{q+1}+\cdots .
\]
By Eq.~\eqref{eq:S8-intertwining-identities}, this is a solution at
level \(q+1\). Its degree is \(q+1\), so it is proportional to the
unique monic lower-degree solution. Hence
\begin{equation}
	\mathcal A_{n+q}^+
	P_{\rm resp}^{(n,q)}
	=
	(2n+q)P_{\rm resp}^{(n,q+1)}.
	\label{eq:S8-Aplus-Presp}
\end{equation}

For the source branch,
\(P_{\rm src}^{(n,q)}=r^{2n+q-1}+\cdots\), so
\[
\mathcal A_{n+q}^+P_{\rm src}^{(n,q)}
=
(q+1)r^{2n+q}+\cdots .
\]
This has the source degree at level \(q+1\). Moreover,
\(\mathcal A_{n+q}^+\) reverses parity, so the result has the parity
of \(P_{\rm src}^{(n,q+1)}\), opposite to that of
\(P_{\rm resp}^{(n,q+1)}\). Thus no response admixture is possible,
and
\begin{equation}
	\mathcal A_{n+q}^+
	P_{\rm src}^{(n,q)}
	=
	(q+1)P_{\rm src}^{(n,q+1)}.
	\label{eq:S8-Aplus-Psrc}
\end{equation}

The lowering relations follow immediately from the monic Gegenbauer
identity
\begin{equation}
	\frac{d}{dr}\mathcal G_m^{(\alpha)}(r)
	=
	m\,\mathcal G_{m-1}^{(\alpha+1)}(r)
	\label{eq:S8-monic-Gegenbauer-derivative}
\end{equation}
and \(\alpha_q+1=\alpha_{q-1}\):
\begin{align}
	\mathcal A_{n+q}^-
	P_{\rm resp}^{(n,q)}
	&=
	qP_{\rm resp}^{(n,q-1)},
	\label{eq:S8-Aminus-Presp}
	\\
	\mathcal A_{n+q}^-
	P_{\rm src}^{(n,q)}
	&=
	(2n+q-1)P_{\rm src}^{(n,q-1)},
	\qquad q\ge1.
	\label{eq:S8-Aminus-Psrc}
\end{align}

Defining
\begin{equation}
	\Phi_{\rm resp}^{(n,q)}
	=
	C_nE_{n+q}P_{\rm resp}^{(n,q)},
	\qquad
	\Phi_{\rm src}^{(n,q)}
	=
	C_nE_{n+q}P_{\rm src}^{(n,q)},
\end{equation}
we obtain
\begin{align}
	J_+\Phi_{\rm resp}^{(n,q)}
	&=
	(2n+q)\Phi_{\rm resp}^{(n,q+1)},
	&
	J_-\Phi_{\rm resp}^{(n,q)}
	&=
	q\Phi_{\rm resp}^{(n,q-1)},
	\label{eq:S8-response-ladder}
	\\
	J_+\Phi_{\rm src}^{(n,q)}
	&=
	(q+1)\Phi_{\rm src}^{(n,q+1)},
	&
	J_-\Phi_{\rm src}^{(n,q)}
	&=
	(2n+q-1)\Phi_{\rm src}^{(n,q-1)},
	\quad q\ge1.
	\label{eq:S8-source-ladder}
\end{align}

At \(q=0\), the response endpoint is the ordinary lowest-weight
state,
\begin{equation}
	J_-\Phi_{\rm resp}^{(n,0)}=0.
	\label{eq:S8-response-lowest-final}
\end{equation}
The source endpoint is not lowest weight. Indeed,
\begin{equation}
	\mathcal A_n^-P_{\rm src}^{(n,0)}
	=
	(2n-1)
	\mathcal G_{2n-2}^{\left(\frac32-n\right)}
	\neq0,
\end{equation}
so \(J_-\Phi_{\rm src}^{(n,0)}\neq0\). Nevertheless,
\(\Phi_{\rm src}^{(n,0)}\) has
\[
J_0\Phi_{\rm src}^{(n,0)}
=
n\Phi_{\rm src}^{(n,0)},
\qquad
\mathcal C\Phi_{\rm src}^{(n,0)}
=
n(n-1)\Phi_{\rm src}^{(n,0)}.
\]
Using
\(\mathcal C=J_0(J_0-1)-J_+J_-\), we therefore obtain directly
\begin{equation}
	J_+J_-\Phi_{\rm src}^{(n,0)}
	=
	\bigl[J_0(J_0-1)-\mathcal C\bigr]
	\Phi_{\rm src}^{(n,0)}
	=
	0.
	\label{eq:S8-source-second-order-null}
\end{equation}

Thus the response branch forms the standard lowest-weight module,
whereas the source branch is a companion ladder whose \(q=0\)
endpoint obeys the second-order relation
\(J_+J_-\Phi_{\rm src}^{(n,0)}=0\) rather than the ordinary
lowest-weight condition.

\suppsubsection{The ingoing resonant polynomial and its source-response decomposition}
\label{subsec:S8-canonical-polynomial-decomposition}

\revstart
At $1\le p\le N$, the solution selected by fixed-$\Delta=p$
continuation of the nonresonant ingoing branch is
\revend

\begin{equation}
	R_{N,p}^{\rm can}(r)
	=
	\chi_N(r)
	{}_2F_1\left(
	p,1-p;1-N;\frac{1-r}{2}
	\right).
	\label{eq:S8-RNp-can}
\end{equation}
To compare it with the descendant family, set
\(p=n\) and \(N=n+q\). Since
\[
\chi_{n+q}(r)
=
s(r)^{-(n+q)}
2^{-(n+q)/2}(r+1)^{n+q},
\]
we may write
\begin{equation}
	R_{n+q,n}^{\rm can}(r)
	=
	s(r)^{-(n+q)}
	P_{\rm can}^{(n,q)}(r),
	\label{eq:S8-Rcan-Pcan-factorization}
\end{equation}
where
\begin{equation}
	P_{\rm can}^{(n,q)}(r)
	=
	2^{-(n+q)/2}(r+1)^{n+q}
	{}_2F_1\left(
	n,1-n;1-n-q;\frac{1-r}{2}
	\right).
	\label{eq:S8-Pcan-definition}
\end{equation}
\revstart
Every hypergeometric symbol in this subsection means the finite sum
\[
G_{N,n}(z)=\sum_{k=0}^{n-1}
\frac{(n)_k(1-n)_k}{(1-N)_k\,k!}z^k,\qquad N=n+q.
\]
All denominators in that sum are nonzero; at \(q=0\) it is
\(T_{n-1}\), not \((1-z)^{-n}\). Thus \(P_{\rm can}^{(n,q)}\)
is a polynomial for every \(q\ge0\). The overall factor
\(2^{-(n+q)/2}\) follows from the normalization of \(\chi_N\).
\revend

As follows from the hypergeometric reduction derived in
Sec.~\ref{sec:S1-radial}, \(R_{n+q,n}^{\rm can}\) satisfies
\(\mathcal D_{n,n+q}R=0\). The identity
\eqref{eq:S8-D-to-L-identity} therefore gives
\begin{equation}
	\mathcal L_{n,q}P_{\rm can}^{(n,q)}=0.
	\label{eq:S8-Pcan-solves-L}
\end{equation}
The two polynomial solutions constructed above form a basis of this
solution space. Hence
\begin{equation}
	P_{\rm can}^{(n,q)}
	=
	W_{\rm src}^{(n,q)}
	P_{\rm src}^{(n,q)}
	+
	W_{\rm resp}^{(n,q)}
	P_{\rm resp}^{(n,q)}.
	\label{eq:S8-Pcan-source-response-decomposition}
\end{equation}
If the notation \(P_q^{(n)}\) is used for the selected polynomial,
then \(P_q^{(n)}\equiv P_{\rm can}^{(n,q)}\), so
Eq.~\eqref{eq:S8-Pcan-source-response-decomposition} is the desired
decomposition.

The basis polynomials are monic, with degrees \(2n+q-1\) and \(q\).
Moreover, their degrees differ by the odd integer \(2n-1\), so they
have opposite parity. Their coefficients are therefore obtained
directly from
\begin{equation}
	W_{\rm src}^{(n,q)}
	=
	[r^{2n+q-1}]P_{\rm can}^{(n,q)},
	\qquad
	W_{\rm resp}^{(n,q)}
	=
	[r^q]P_{\rm can}^{(n,q)},
	\label{eq:S8-Wsrc-Wresp-extraction}
\end{equation}
where \([r^m]\) denotes the coefficient of \(r^m\).

Multiplying
Eq.~\eqref{eq:S8-Pcan-source-response-decomposition} by
\(s(r)^{-(n+q)}\) gives
\begin{equation}
	R_{n+q,n}^{\rm can}
	=
	W_{\rm src}^{(n,q)}
	R_{\rm src}^{(n,q)}
	+
	W_{\rm resp}^{(n,q)}
	R_{\rm resp}^{(n,q)}.
	\label{eq:S8-Rcan-source-response-decomposition}
\end{equation}
\revstart
The decomposition coefficients have the closed forms, with \(N=n+q\),
\begin{align}
W_{\rm src}^{(n,q)}&=(-1)^{n-1}2^{1-n-N/2}
\frac{(2n-2)!(N-n)!}{(n-1)!(N-1)!},
\tag{S158a}\label{eq:S8-Wsrc-closed}\\
W_{\rm resp}^{(n,q)}&=2^{n-N/2-1}
\frac{(n-1)!(N+n-1)!}{(2n-1)!(N-1)!}.
\tag{S158b}\label{eq:S8-Wresp-closed}
\end{align}
These are exactly \(A_0,B_0\) in the monic parity basis. An independent
finite-polynomial derivation uses \(P_{\rm can}(-1)=0\) and
\(P_{\rm can}(1)=2^{N/2}\). The two Gegenbauer polynomials have
opposite parity, so
\[
W_{\rm src}P_{\rm src}(1)=W_{\rm resp}P_{\rm resp}(1)=2^{N/2-1}.
\]
Now \(\mathcal G_m^{(\alpha)}(1)=(2\alpha)_m/[2^m(\alpha)_m]\),
with \(\alpha=1/2-N\); evaluating these finite products gives
Eqs.~\eqref{eq:S8-Wsrc-closed}--\eqref{eq:S8-Wresp-closed}.
They are nonzero for every \(n\ge1,q\ge0\).
Furthermore, the continued boundary branches obey
\(R_1=2^{-N/2}R_{\rm src}^{(n,q)}\) and
\(R_2=2^{-N/2}R_{\rm resp}^{(n,q)}\).

At \(q=0\), the monic basis simplifies to
\begin{equation}
P_{\rm resp}^{(n,0)}=1,\qquad
P_{\rm src}^{(n,0)}=(2n-1)\int_0^r(t^2-1)^{n-1}dt.
\tag{S158c}\label{eq:S8-endpoint-source-integral}
\end{equation}
Thus \(W_{\rm src}^{(n,0)}=(-1)^{n-1}2^{1-3n/2}
\binom{2n-2}{n-1}\) and \(W_{\rm resp}^{(n,0)}=2^{n/2-1}\).
For \(n=1\), \(R_{1,1}^{\rm can}=(r+1)/(\sqrt2s)\), whereas
\(R_2=1/(\sqrt2s)\). For \(n=2\),
\[
R_{2,2}^{\rm can}=-\frac12\frac{r^3-3r}{r^2-1}
+\frac1{r^2-1},\qquad R_2=\frac1{2(r^2-1)}.
\]
Finally, \(J_-\Phi_{\rm can}^{(n,0)}
=W_{\rm src}^{(n,0)}(2n-1)e^{-(n-1)t}s^{n-1}\ne0\), while
\(J_-\Phi_{\rm resp}^{(n,0)}=0\). The response ladder therefore
generates the response component, not the full canonical mode,
including at the endpoint.
\revend

\suppsubsection{Direct radial check for the selected resonant branch}
\label{subsec:S8-direct-radial-check-canonical}

The radial-to-hypergeometric reduction was derived in
Sec.~\ref{sec:S1-radial}. \revstart
For $1\le p\le N$, let
\[
F(z)=\sum_{k=0}^{p-1}\frac{(p)_k(1-p)_k}{(1-N)_k\,k!}z^k.
\]
\revend

the parameters \(a=p\), \(b=1-p\), and \(c=1-N\) give
\(a+b+1=2\) and \(-ab=p(p-1)\). Hence the hypergeometric equation is
precisely the residual radial equation with
\(\Delta=p\) and \(\mu=N\). Therefore
\begin{equation}
	\mathcal D_{p,N}
	\left[
	\chi_N(r)
	{}_2F_1\left(
	p,1-p;1-N;\frac{1-r}{2}
	\right)
	\right]
	=0.
	\label{eq:S8-RNp-can-solves-D}
\end{equation}
For \(p=n\) and \(N=n+q\), this confirms
\(\mathcal D_{n,n+q}R_{n+q,n}^{\rm can}=0\).

\suppsubsection{Interpretation}
\label{subsec:S8-interpretation-two-branches}

The \(SL(2,\mathbb R)\) construction reorganizes the pole-skipping
points with fixed \(\Delta=n\) into the descendant sequence
\(\mu=n+q\), \(q\in\mathbb Z_{\ge0}\). At each level, the auxiliary
equation has a degree-\(q\) response polynomial and a
degree-\((2n+q-1)\) source polynomial, producing the boundary
behaviors \(r^{-n}\) and \(r^{n-1}\), respectively.

The ordinary lowest-weight module generated from
\(\Phi_{\rm resp}^{(n,0)}\) therefore singles out the response branch,
but does not exhaust the full two-dimensional radial solution space.
Rather, the ingoing resonant solution is the linear combination
\begin{equation}
	R_{n+q,n}^{\rm can}
	=
	W_{\rm src}^{(n,q)}R_{\rm src}^{(n,q)}
	+
	W_{\rm resp}^{(n,q)}R_{\rm resp}^{(n,q)}.
	\label{eq:S8-interpretation-canonical-decomposition}
\end{equation}
Thus the response ladder is one component of the resonant solution, while the higher-degree source branch is its independent companion. This gives the relation between the lowest-weight construction and the ingoing resonant solution.

\suppsection{Static HSC diagnostic: smoothness and linear independence}
\label{sec:S9-HSC-diagnostic}

In this section we use the static sector of a holographic
superconductor with a \((3+1)\)-dimensional boundary theory as a
diagnostic example. The purpose is not to identify a new
pole-skipping point in this sector. Rather, the example shows that the
criterion of the main text requires both horizon residual smoothness
and linear independence of the two boundary-normalized branches.

We begin with the Abelian--Higgs matter-sector action in the
five-dimensional bulk,
\begin{equation}
	S
	=
	\int d^5x\sqrt{-g}
	\left[
	-\frac14 F_{\mu\nu}F^{\mu\nu}
	-|D\Psi|^2
	-m^2|\Psi|^2
	\right],
	\qquad
	D_\mu=\nabla_\mu-ieA_\mu .
	\label{eq:S9-AH-action}
\end{equation}
In the static near-critical sector, the scalar equation reduces to a
second-order ordinary differential equation. In the dimensionless
coordinate \(x\), with the AdS boundary at \(x=0\) and the black-hole
horizon at \(x=1\), it takes the form
\begin{equation}
	-\Psi''(x)
	+
	\frac{x^2+1}{x(1-x^2)}\Psi'(x)
	+
	\left[
	\frac{\Delta(\Delta-4)}{4x^2(1-x^2)}
	-
	\frac{\lambda^2}{4x(x+1)^2}
	\right]\Psi(x)
	=
	0,
	\label{eq:S9-HSC-static-equation}
\end{equation}
where \(\lambda\) is the dimensionless electric-field parameter and
\(m^2=\Delta(\Delta-4)\).

\suppsubsection{Boundary branches}
\label{subsec:S9-boundary-branches}

Near the AdS boundary \(x=0\), the Frobenius substitution
\(\Psi\sim x^\rho\) gives
\begin{equation}
	\rho_+=\frac{\Delta}{2},
	\qquad
	\rho_-=\frac{4-\Delta}{2},
	\label{eq:S9-boundary-roots}
\end{equation}
and hence
\begin{equation}
	\Psi_+(x)\sim x^{\Delta/2},
	\qquad
	\Psi_-(x)\sim x^{(4-\Delta)/2}.
	\label{eq:S9-two-boundary-branches}
\end{equation}
In the standard quantization used here, \(\Psi_+\) and \(\Psi_-\)
denote the response and source branches, respectively.

We factor the regular contribution at \(x=-1\) and write
\begin{align}
	\Psi_+(x)
	&=
	x^{\Delta/2}(1+x)^{-\lambda/2}
	\sum_{k=0}^{\infty}d_k^{(+)}x^k,
	\nonumber\\
	\Psi_-(x)
	&=
	x^{(4-\Delta)/2}(1+x)^{-\lambda/2}
	\sum_{k=0}^{\infty}d_k^{(-)}x^k,
	\label{eq:S9-boundary-ansatze}
\end{align}
with
\[
d_0^{(\pm)}=1,
\qquad
d_{-1}^{(\pm)}=0.
\]
For the response branch, substitution into
Eq.~\eqref{eq:S9-HSC-static-equation} gives the un-divided recurrence
\begin{equation}
	(k+1)(k+\Delta-1)d_{k+1}^{(+)}
	-
	\frac{\lambda}{2}
	\left(
	2k+\Delta-1-\frac{\lambda}{2}
	\right)d_k^{(+)}
	-
	\left(
	k-1+\frac{\Delta}{2}-\frac{\lambda}{2}
	\right)^2
	d_{k-1}^{(+)}
	=0,
	\qquad k\ge0.
	\label{eq:S9-response-recurrence}
\end{equation}
The source recurrence follows from the exact mirror transformation
\begin{equation}
	\Delta\longmapsto4-\Delta
	\label{eq:S9-boundary-mirror-map}
\end{equation}
applied to Eq.~\eqref{eq:S9-response-recurrence}.

The boundary exponent difference is
\begin{equation}
	\rho_+-\rho_-=\Delta-2.
	\label{eq:S9-boundary-exponent-difference}
\end{equation}
Thus \(\Delta\in\mathbb Z\) gives a resonant boundary Frobenius
problem, including the coincident-root case \(\Delta=2\). At these
values, a nominal second branch may become logarithmic or collapse
onto the first one. We therefore require the residual series in
Eq.~\eqref{eq:S9-boundary-ansatze} to define log-free
\(C^\infty\) functions at \(x=0\).
\revstart
At integer resonance the displayed source normalization
\(d_0^{(-)}=1\) is only nominal: if its log-free compatibility
condition fails, the admissible log-free representative instead begins
at the first permitted higher order, as explained in
Sec.~\ref{subsec:S9-branch-collapse}.
\revend

\suppsubsection{Exterior horizon-smoothness loci}
\label{subsec:S9-three-loci}

\begin{figure}[t]
	\centering
	\includegraphics[width=0.70\linewidth]{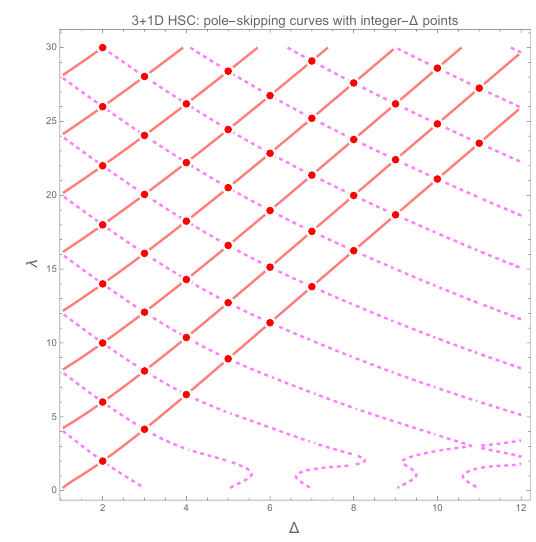}
	\caption{\revstart
		Static HSC exterior horizon-smoothness loci.
		The red solid curves denote \(L_+(\Delta,\lambda)=0\) for the
		response representative; the pink dashed curves denote
		\(L_-(\Delta,\lambda)=0\) for the source representative.
		The marked integer-\(\Delta\) coincidences illustrate apparent
		two-branch candidates. When the resonant log-free source branch
		collapses, it is proportional to the response branch, as shown
		by the Frobenius compatibility calculation below. Irrespective
		of the numerical marker precision, two horizon-smooth solutions
		in this static sector cannot be independent, by Abel's identity.
		\revend}
	\label{fig:S9-HSC-diagnostic}
\end{figure}

\revstart
We now restrict attention to the two exterior boundary branches.
Let \(L_+(\Delta,\lambda)\) and \(L_-(\Delta,\lambda)\) denote the
coefficients of the horizon logarithm obtained by continuing the
response and source representatives, respectively. At the static
horizon, the indicial root is repeated and equal to zero; consequently,
a local logarithmic contribution has the form
\(L_\pm H(1-x)\log(1-x)\), with \(H\) analytic and nonzero at the
horizon. The respective exterior smoothness conditions are
\revend
\begin{align}
	L_+(\Delta,\lambda)&=0,
	\label{eq:S9-Lresp}
	\\
	L_-(\Delta,\lambda)&=0.
	\label{eq:S9-Lsrc}
\end{align}
\revstart
Figure~\ref{fig:S9-HSC-diagnostic} displays the response locus in red
and the source locus in pink dashed lines. The marked coincidences at
integer \(\Delta\) motivate the boundary Frobenius compatibility
analysis in Sec.~\ref{subsec:S9-branch-collapse}. The figure is a
finite-truncation diagnostic, not an independent exact certificate of
every displayed intersection. The linear-independence obstruction
below follows analytically from Abel's identity, without an
exterior--interior matching calculation.
\revend

\setcounter{equation}{180}

\suppsubsection{Genuine pole-skipping criterion}
\label{subsec:S9-genuine-criterion}

Whenever the two boundary-normalized solutions are linearly
independent, a general solution may be written as
\begin{equation}
	\Psi(x)=A\Psi_-(x)+B\Psi_+(x).
	\label{eq:S9-general-boundary-solution}
\end{equation}
Because the horizon logarithmic coefficient is linear in the solution,
the no-log condition is
\begin{equation}
	A L_-(\Delta,\lambda)
	+
	B L_+(\Delta,\lambda)
	=0.
	\label{eq:S9-horizon-nolog-condition}
\end{equation}
At a generic point this determines
\begin{equation}
	\frac{B}{A}
	=
	-\frac{L_-(\Delta,\lambda)}
	{L_+(\Delta,\lambda)}.
	\label{eq:S9-ratio-generic}
\end{equation}
If
\begin{equation}
	L_+(\Delta,\lambda)
	=
	L_-(\Delta,\lambda)
	=
	0,
	\label{eq:S9-two-L-zero}
\end{equation}
the horizon condition becomes \(0=0\) and no longer fixes \(B/A\).

\revstart
These vanishing conditions are only a candidate pole-zero degeneracy;
they do not by themselves prove a path-dependent Green function.
For a genuine two-branch interpretation, the boundary-admissible
solutions must first remain independent. In the static equation,
Abel's identity gives the exact stronger obstruction
\begin{equation}
W[\Psi_+,\Psi_-](x)=C\frac{x}{1-x^2}.
\tag{S184a}\label{eq:S9-static-Abel-obstruction}
\end{equation}
Both solutions being \(C^\infty\) at \(x=1\) would make their
Wronskian bounded; Eq.~\eqref{eq:S9-static-Abel-obstruction} then
forces \(C=0\). Thus the horizon-smooth space is at most
one-dimensional, in agreement with its repeated zero indicial root.
Any simultaneous smooth representatives in this static sector fail
the required independence condition. This statement holds
independently of the plotted intersection positions.
\revend
 By Abel's
identity, it is enough to test their Wronskian at any ordinary point
\(x_0\in(0,1)\):
\begin{equation}
	W[\Psi_+,\Psi_-](x_0)
	=
	\Psi_+(x_0)\Psi_-'(x_0)
	-
	\Psi_+'(x_0)\Psi_-(x_0).
\end{equation}
Thus
\begin{equation}
	\boxed{
		\revstart
\text{necessary two-branch conditions:}
		\quad
\revend
		L_+=L_-=0
		\quad\text{and}\quad
		W[\Psi_+,\Psi_-]\neq0 .
	}
	\label{eq:S9-genuine-criterion}
\end{equation}

\revstart
The apparent simultaneous-smoothness candidates in the displayed
static HSC diagram therefore fail the independence test.

\revend

\suppsubsection{Branch collapse at integer \(\Delta\)}
\label{subsec:S9-branch-collapse}

\revstart
Figure~\ref{fig:S9-HSC-diagnostic} marks red/pink apparent
coincidences at nominally integer \(\Delta\), motivating a
boundary-resonance test. Integer resonance
alone, however, does not automatically
imply branch collapse; one must examine the Frobenius compatibility
condition.
\revend

Let
\[
\Delta=p\ge3,
\qquad
m=p-2,
\]
so that the boundary exponents differ by \(m\). The un-divided
recurrence for the nominal source branch is
\begin{align}
	&(k+1)(k+3-p)d_{k+1}^{(-)}
	\nonumber\\
	&\quad
	-
	\frac{\lambda}{2}
	\left(
	2k+3-p-\frac{\lambda}{2}
	\right)d_k^{(-)}
	-
	\left(
	k+1-\frac{p}{2}-\frac{\lambda}{2}
	\right)^2
	d_{k-1}^{(-)}
	=0 .
	\label{eq:S9-source-recurrence-integer}
\end{align}
For \(0\le k<m-1\), this recurrence determines
\(d_1^{(-)},\ldots,d_{m-1}^{(-)}\) in terms of \(d_0^{(-)}\).
At the resonant step \(k=m-1=p-3\), the coefficient of
\(d_m^{(-)}\) vanishes, leaving a compatibility condition of the form
\begin{equation}
	\mathcal K_m(\lambda)\,d_0^{(-)}=0,
	\label{eq:S9-compatibility-obstruction}
\end{equation}
where \(\mathcal K_m(\lambda)\) is determined by the preceding
recurrence steps.

If
\begin{equation}
	\mathcal K_m(\lambda)\neq0,
	\label{eq:S9-obstruction-nonzero}
\end{equation}
the log-free prescription forces \(d_0^{(-)}=0\). The preceding
nonresonant recurrence steps then successively imply
\[
d_1^{(-)}
=
\cdots
=
d_{m-1}^{(-)}
=
0,
\]
while \(d_m^{(-)}\) becomes the first free coefficient. Consequently,
\begin{align}
	\Psi_-(x)
	&=
	x^{(4-p)/2}(1+x)^{-\lambda/2}
	\left[
	d_m^{(-)}x^m+O(x^{m+1})
	\right]
	\nonumber\\
	&=
	x^{p/2}(1+x)^{-\lambda/2}
	\left[
	d_m^{(-)}+O(x)
	\right].
	\label{eq:S9-general-branch-collapse}
\end{align}
The nominal source branch therefore acquires the response exponent and
collapses onto the unique log-free response branch. Hence
\begin{equation}
	W[\Psi_+,\Psi_-]=0.
	\label{eq:S9-collapse-wronskian}
\end{equation}

For example, at \(p=3\) one has \(m=1\), and the resonant first step is
\begin{equation}
	\frac{\lambda^2}{4}d_0^{(-)}=0.
	\label{eq:S9-Delta3-first-step}
\end{equation}
Thus, for \(\lambda\neq0\),
\[
d_0^{(-)}=0,
\qquad
\Psi_-(x)
=
x^{3/2}(1+x)^{-\lambda/2}
\left[d_1^{(-)}+O(x)\right],
\]
which is the response behavior. The case \(p=1\) follows by the mirror
map \(p\mapsto4-p\).

At \(p=2\), the two indicial roots coincide:
\[
\rho_+=\rho_-=1.
\]
The local basis consists of one log-free branch and one logarithmic
branch, so removing the logarithmic solution again leaves only one
boundary-smooth representative.

\revstart
The preceding boundary-collapse calculation is conditional on
\(\mathcal K_m(\lambda)\ne0\). Integer resonance alone does not prove
that condition, and no plot-by-plot exact obstruction values are
asserted here. At points where \(\mathcal K_m=0\), independent
boundary log-free branches may exist, but they still cannot both be
smooth at the static horizon by Eq.~\eqref{eq:S9-static-Abel-obstruction}.
Thus the exclusion of a genuine two-branch smoothness intersection
does not require assuming a nonzero obstruction at every plotted point.

\setcounter{equation}{194}

\revend
\setcounter{equation}{196}

\suppsubsection{Conclusion of the diagnostic}
\label{subsec:S9-HSC-conclusion}

\revstart
Any simultaneous horizon-smooth representatives in this static HSC
equation are dependent. Therefore they cannot constitute a genuine
two-branch pole-zero intersection. The exact conclusion is
\revend

\begin{equation}
	\boxed{
		\revstart
		\text{simultaneous static horizon smoothness}
		\Longrightarrow
		W[\Psi_+,\Psi_-]=0
		\revend
	}.
	\label{eq:S9-HSC-diagnostic-conclusion}
\end{equation}
 \revstart
 This example illustrates the independence condition used
 in the JT/AdS$_2$ analysis: two horizon-smooth
 boundary representatives must remain linearly
 independent to constitute a genuine two-branch
 pole-zero intersection. When the representatives
 are proportional, they span only a one-dimensional
 space of bulk solutions. Consequently, a formal
 simultaneous vanishing of their horizon logarithmic
 coefficients does not establish a genuine
 pole-skipping ambiguity.
\revend
\end{document}